\documentclass[aip,amsmath,amssymb,reprint]{revtex4-1}
\usepackage{comment}
\usepackage[caption = false]{subfig}
\usepackage{graphicx}
\usepackage{dcolumn}
\usepackage{bm}
\usepackage[utf8]{inputenc}
\usepackage[T1]{fontenc}
\usepackage{mathptmx}
\usepackage{etoolbox}
\usepackage{siunitx}
\DeclareSIUnit\angstrom{\text {Å}}
\usepackage{mathtools}
\usepackage{longtable}
\usepackage{xr}
\usepackage{subfiles}
\usepackage{color,soul}
\usepackage[normalem]{ulem}
\usepackage[export]{adjustbox}

\makeatletter
\def\@email#1#2{%
 \endgroup
 \patchcmd{\titleblock@produce}
  {\frontmatter@RRAPformat}
  {\frontmatter@RRAPformat{\produce@RRAP{*#1\href{mailto:#2}{#2}}}\frontmatter@RRAPformat}
  {}{}
}%
\makeatother

\begin{document}
\preprint{AIP/123-QED}
\title{Phaseless auxiliary field quantum Monte Carlo with projector-augmented wave method for solids}

\author{Amir Taheridehkordi}
\email{amir.taheridehkordi@univie.ac.at}
\affiliation{University of Vienna, Faculty of Physics, Kolingasse 14-16, A-1090 Vienna, Austria}
\author{Martin Schlipf}
\affiliation{VASP Software GmbH, Sensengasse 8, A-1090 Vienna, Austria}
\author{Zoran Sukurma}
\affiliation{University of Vienna, Faculty of Physics \& Vienna Doctoral School in Physics,  Boltzmanngasse 5, A-1090 Vienna, Austria}
\author{Moritz Humer}
\affiliation{University of Vienna, Faculty of Physics  \& Vienna Doctoral School in Physics,  Boltzmanngasse 5, A-1090 Vienna, Austria}
\author{Andreas Gr\"uneis}
\affiliation{Institute for Theoretical Physics, TU Wien, Wiedner Hauptstraße 8-10/136, 1040 Vienna, Austria}
\author{Georg Kresse}
\affiliation{University of Vienna, Faculty of Physics, Kolingasse 14-16, A-1090 Vienna, Austria}
\affiliation{VASP Software GmbH, Sensengasse 8, A-1090 Vienna, Austria}

\date{\today}
\begin{abstract}
We implement the phaseless auxiliary field quantum Monte Carlo method using the plane-wave based projector augmented wave method and explore the accuracy and the feasibility of applying our implementation to solids. 
We use a singular value decomposition to compress the two-body Hamiltonian and thus reduce the computational cost. 
Consistent correlation energies from the primitive-cell sampling and the corresponding supercell calculations numerically verify our implementation. 
We calculate the equation of state for diamond and the correlation energies for a range of prototypical solid materials. 
A down-sampling technique along with natural orbitals accelerates the convergence with respect to the number of orbitals and crystal momentum points. 
We illustrate the competitiveness of our implementation in accuracy and computational cost for dense crystal momentum point meshes comparing to a well-established quantum-chemistry approach, the coupled-cluster ansatz including singles, doubles and perturbative triple particle-hole excitation operators.
\end{abstract}

\maketitle

\section{Introduction} \label{Sec:Intro}

One of the most challenging tasks in solid-state physics is to solve the many-electron Schr\"odinger equation.
For this purpose effective one-electron methods based on density functional theory \cite{Kohn:1965,Kohn:1999} (DFT)
are particularly successful.
Introducing an exchange-correlation density functional, DFT replaces the electron-electron interaction with an effective one-electron potential.
Thus, the interacting many-electron problem is reduced to a set of one-electron equations, which can be solved self-consistently
in a representation defined by the employed basis set.
For solids, plane waves are the most popular basis functions because they
exhibit a favorable scaling with system size and are independent of atom positions and species.
Moreover, their convergence is systematically controlled by the plane-wave energy cutoff.  
 
Projector augmented waves (PAWs) are an efficient way to describe the all-electron DFT orbitals.\cite{Bloechl1994,Kresse1999}
In this method, one constructs the all-electron orbital replacing a certain fraction of a pseudo-orbital with 
a localized function in the vicinity of the ions.
The specific fraction depends on the overlap of the pseudo-orbital with so-called \emph{projectors}.
The important point is that the PAW method yields essentially all-electron precision but the plane-wave energy cutoff corresponds to the smoother pseudo-orbital. 
The high precision of the PAW methods has been demonstrated for density functional theory methods for both, small molecules \cite{Paier:2005} as well as solids.\cite{lejaeghere2016} Recently the evaluations were also extended to many-body calculations for small molecules,\cite{Moritz:2022} demonstrating that PAW potentials can reach chemical accuracy ($<$~1 kcal/mol). 

The main issue of DFT is the choice of the exchange-correlation functional.
Since calculating the exact exchange-correlation functional has the same complexity as solving the Schr\"odinger equation,
in practice, one needs to approximate the functional.
These approximations lead to inaccurate results especially in strongly-correlated systems.\cite{Burke:2012} 
Other weaknesses of DFT include the description of thermochemical properties \cite{Curtiss:1997,Paier:2007} and van der Waals interactions.\cite{dobson:2001prediction} 

Alternatively, one can find the ground state of the Schr\"odinger equation explicitly.
Quantum-chemical wavefunction based methods are limited to small-sized systems because of the adverse scaling of their computational cost.
For example, the computational cost of the coupled-cluster ansatz using single, double and perturbative triple particle–hole excitation operators \cite{RAGHAVACHARI:1989,Bartlett:2007,Gruneis:2015,Gruber:2018} (CCSD(T))
scales with the seventh power of the system size and configuration interaction methods\cite{szabo2012modern,cramer:2013,Vogiatzis_FCI:2017} scale exponentially.
Although this adverse scaling can be reduced by adopting local correlation methods, it is not guaranteed that local correlation methods entirely avoid uncontrolled errors. 
For example, recent work on large molecules yielded conflicting results for localized coupled-cluster methods and diffusion Monte Carlo (DMC) methods---the reason for this inconsistency not yet been known. \cite{al2021interactions,nagy2018optimization,nagy2019approaching,zalesny2011linear}
Furthermore, for densely packed 3D solids, it is not clear whether local methods can accelerate the calculations in the same way as they do for more open low-dimensional structures.
Quantum Monte Carlo\cite{kalos:1974,ceperley:1995,blankenbecler:1981} (QMC) methods
also overcome adverse scaling with a cubic to quartic increase of the computational cost with system size.
However, DMC\cite{Anderson:1976,Ceperley:1977,foulkes:2001} requires local potentials and, hence, has  not yet been
combined with the PAW method, since the PAW methods relies on non-local projectors. Even, when non-local pseudopotentials are used in DMC, approximations must be made.\cite{Casula:2006,Casula:2010,Anderson:2021}
Variational Monte Carlo\cite{McMillan:1965}
cannot typically reach chemical accuracy.\cite{Nemec:2010} 
Full-configuration interaction QMC\cite{Booth:2009,Cleland:2010,Ghanem:2020} (FCIQMC) and semi-stochastic heat-bath configuration interaction\cite{sharma2017semistochastic,yao2020almost} as a variant of selected configuration interaction methods\cite{holmes2016heat,dash2018perturbatively} are very accurate but still contain terms scaling weakly exponentially with system size.

In the auxiliary-field quantum Monte Carlo (AFQMC) method,\cite{Sorella:1989,Zhang:1995,zhang:1997constrained,Baer:1998,Purwanto:2004,motta_review:2018,Purwanto_si:2009,motta_redmat:2017,motta_comm:2018,Purwanto_exc:2009,Ma_exc:2013,Al-Saidi_dt_nw:2006,Suewattana:2007,esler:2008,Motta:2019}
one represents the ground-state wavefunction as an ensemble of Slater determinants. AFQMC  utilizes two key ideas:
First, similar to other projective QMC approaches, it constructs the ground-state wavefunction by repeated application of an infinitesimal imaginary-time propagator to a trial wavefunction. 
Second, the Hubbard-Stratonovich transformation \cite{Stratonovich:1957,hubbard:1959} reduces the interacting many-body problem to a high-dimensional integral of one-body operators using auxiliary fields. 
In real materials, one needs to carefully control the phase of the ensemble of walkers to ensure numeric stability.
This so-called \emph{phase problem} is a generalization of the sign problem \cite{loh:1990sign} and is mitigated by the phaseless approximation in AFQMC (ph-AFQMC).\cite{Zhang:2003}

ph-AFQMC has been applied to a wide range of systems from the Hubbard model \cite{Hubbard:1963,Zheng:2017} to atoms and molecular systems.\cite{morales:2019,Shee:2019,Kiel:2020}
Using Gaussian-type orbitals, ph-AFQMC was compared to DMC in 
solids\cite{malone:2020} and applied to solid NiO.\cite{Zhang_no:2018}
Zhang \emph{et al.} implemented ph-AFQMC for plane waves using norm-conserving 
pseudopotentials.\cite{Zhang:2003}
Furthermore, ph-AFQMC has been used to study the pressure-induced transition in silicon \cite{Purwanto_si:2009} and to compute benchmark charge densities for solids.\cite{chen:2021}
Algorithmic improvements reduce the computational cost with down-sampling and frozen 
orbitals\cite{Purwanto:2013,Ma:2015} and generalize ph-AFQMC for optimized 
norm-conserving pseudopotentials.\cite{ma:2017}
All these successes motivate us to combine ph-AFQMC with the PAW method to expand the limits of \emph{ab initio} calculations.

In this work, we implement ph-AFQMC using the plane-wave based PAW method.
Our aim is to explore whether ph-AFQMC for solids is feasible and what kind of accuracy can be expected for simple prototypical solids.
To achieve this goal, we compare ph-AFQMC with widely-used quantum-chemistry methods---second-order M\o ller-Plesset perturbation theory \cite{Moller:1934,Marsman:2009,Gruneis:2010} (MP2), and coupled-cluster calculations at the level of CCSD and CCSD(T).
The numerical setup is intentionally reduced so that one can compare all these methods. Therefore, the results in this work cannot be compared directly with the experiment.
We utilize a singular value decomposition (SVD) \cite{strang1993introduction} to reduce the computational cost.
In general, we find that ph-AFQMC yields slightly larger absolute correlation energies than CCSD(T). 
However, the correlation-energy difference between ph-AFQMC and CCSD(T) is one order of magnitude smaller than the one between ph-AFQMC and MP2.
Calculations are performed for both primitive cells and supercells, with excellent agreement between them, validating the implementation for primitive cells using $\mathbf k$ points.

In addition, we demonstrate down-sampling strategies to converge the results with respect to the employed $\mathbf k$ points and the number of unoccupied bands. 
These strategies facilitate computing the diamond correlation energy approximately in the complete basis-set limit. 
We clearly show that correlation energies of similar accuracy as CCSD(T) can be achieved. Finally, we compare the computational cost of the code to CCSD(T) as a function of the number of $\mathbf k$ points. 
The present implementation is in Python\cite{Rossum:2009} and far from being fully optimized.
Still, the better scaling of ph-AFQMC leads to faster execution times than using a Fortran\cite{metcalf1999FORTRAN} CCSD(T) code\cite{Grueneis:2011,Gruneis:2015} for dense $\mathbf k$-point grids.

The remainder of this paper is structured as follows:
In Sec.~\ref{Sec:Method}, we 
introduce the ph-AFQMC method and show how the application of an SVD reduces the computational cost.
In Sec.~\ref{Sec:result}, we summarize our results for lattice constants and correlation energies.
We compare ph-AFQMC with other quantum-chemistry methods and specifically show that it is competitive in accuracy with CCSD(T). 
Finally, we conclude in Sec.~\ref{Sec:conclusion}.

\section{Method} \label{Sec:Method}

This section describes the required tasks to obtain the ph-AFQMC ground-state energy starting from the Born-Oppenheimer Hamiltonian. 
First, we reduce the computational cost applying an SVD.
Second, the Hamiltonian is written in a mean-field subtracted form to reduce the variance of the ground-state energy. 
Third, we discuss the update of the Slater determinants and how the ground-state energy is calculated. These steps incur the largest computational cost. 
Finally, we lay out the general procedure used to obtain the numeric results.

\subsection{Hamiltonian}\label{sec:hamil}

We split the electronic Born-Oppenheimer Hamiltonian
\begin{eqnarray}\label{E:Hamil_tot_A}
\hat H =  \hat H_1' + \hat H_2',
\end{eqnarray}
into a single-particle operator
\begin{eqnarray}\label{E:H_1}
\hat H_1' = \sum_{\mathbf k} \sum_{pq} h_{pq} (\mathbf k) \hat a^\dagger_{p\mathbf k} \hat a_{q\mathbf k},
\end{eqnarray}
and a two-particle operator 
\begin{eqnarray}\label{E:H_2}
\hat H_2' = \frac{1}{2} \sum_{\mathbf q \mathbf G}  \hat L'_{\mathbf{qG}} \hat L'^\dagger_{\mathbf{qG}}.
\end{eqnarray}
In Eq.~(\ref{E:H_1}), $\hat a_{p\mathbf k}$ ($\hat a^\dagger_{p\mathbf k}$) is the fermionic annihilation (creation) operator, $p$ and $q$ are band indices and $\mathbf k$ is the crystal momentum. 
In Eq.~(\ref{E:H_2}), $\mathbf q$ and $\mathbf G$ are the transferred crystal momentum and a reciprocal lattice vector, respectively. 
For more details we refer the reader to Appendix \ref{appendix:hamil}, where we describe the contributions to the matrix elements $h_{pq}$ 
and how to rewrite the two-body part in terms of single-particle operators 
$\hat L'_{\mathbf{qG}}$ as expressed in Eq.~(\ref{E:H_2}). 
We neglect the spin indices for the sake of simplicity.

\subsection{Application of SVD}

\begin{table}
\caption{\label{tab:mts_size} Dimension of matrices introduced in Eqs.~(\ref{E:svd_gen}) and (\ref{E:compressed_L}). Applying an SVD and truncating the singular values reduces one dimension of $L'_{\mathbf q }$ from the number of plane waves $N_{\mathbf G}$ to the number of truncated singular values
$n_{\mathbf q g} \le N_{\rm s} \leq \min( N_{\rm b}^2, N_{\mathbf G})$ where $N_{\mathrm s}$ is the number of singular values and $N_{\mathrm b}$ is the number of orbitals.
}
\begin{ruledtabular}
\begin{tabular} {cccc}
\multicolumn{2}{c}{in Eq.~\eqref{E:svd_gen}} & \multicolumn{2}{c}{in Eq.~\eqref{E:compressed_L}} \\
matrix & dimension  & matrix & dimension \\
\hline
$U$  & $N_{\rm b}^2 \times N_{\rm s}$  &
$U'$ & $N_{\rm b}^2 \times n_{\mathbf q g}$
\\
$\Sigma$  & $N_{\rm s} \times N_{\rm s}$ &
$\Sigma'$ & $n_{\mathbf q g} \times n_{\mathbf q g}$
\\
$V^\dagger$  & $N_{\rm s} \times N_{\mathbf G}$ &
\\
$L'_{\mathbf q }$          & $N_{\rm b}^2 \times N_{\mathbf G}$ &
$\mathcal L'_{\mathbf q }$ & $N_{\rm b}^2 \times n_{\mathbf q g}$
\\
\end{tabular}
\end{ruledtabular}
\end{table}
For each $\mathbf q$, we represent the $\hat L'_{\mathbf q \mathbf G}$ operators as a matrix $L'_{\mathbf q}$ (see Eqs.~(\ref{E:def_rho}), (\ref{E:A_Q_G}) and (\ref{E:L_mat_rep})).
To reduce the computational cost, we reduce the size of these matrices by an SVD 
\begin{eqnarray}\label{E:svd_gen}
L'_{\mathbf q } = U \Sigma V^\dagger.
\end{eqnarray}
$\Sigma$ is a diagonal matrix containing $N_s$ nonzero singular values. $U$ and $V$ are semi-unitary matrices, i.e., $U^\dagger U = VV^\dagger = \mathbb{I}$, where $\mathbb{I}$ is the identity matrix. 
The dimension of $U$ and $V^\dagger$ are $N_{\rm b}^2 \times N_{\rm s}$ and $N_{\rm s} \times N_{\mathbf G}$, respectively, where $N_b$ is the number of orbitals. 
The number of singular values $N_{\rm s} \leq \min( N_{\rm b}^2, N_{\mathbf G})$ is reduced to $n_{\mathbf q g} \le N_{\rm s}$ by neglecting values smaller than a particular threshold.
Empirically, we find that $n_{\mathbf q g}$ is an order of magnitude smaller than $N_{\mathbf G}$.
In passing we note that a similar approach was used for reducing the computational
cost of integral calculations in periodic coupled-cluster theory with a plane-wave basis.\cite{Hummel:2017}

We then approximate the two-body part of the Hamiltonian by
\begin{eqnarray}\label{E:H_2_svd}
\hat H_2' = \frac{1}{2} \sum_{\mathbf q}\sum_{g=1}^{n_{\mathbf q g}}  \hat {\mathcal L}'_{\mathbf{q}g}  \hat {\mathcal L}'^\dagger_{\mathbf{q}g},
\end{eqnarray}
where in the matrix representation we have
\begin{eqnarray}\label{E:compressed_L}
{\mathcal L'_{\mathbf q }} = U' \Sigma'. 
\end{eqnarray}
Here, the dimension of $U'$ and $\Sigma'$ are $N_{\rm b}^2 \times n_{\mathbf q g}$ and $n_{\mathbf q g} \times n_{\mathbf q g}$, respectively. Table.~\ref{tab:mts_size} summarizes the matrix dimensions in Eqs.~(\ref{E:svd_gen}) and (\ref{E:compressed_L}). 

Introducing 
\begin{align}\label{E:def_e}
\begin{split}
 \hat {\mathcal L}^{({\rm e})}_{{\mathbf{q}g}} =  \frac{1}{2}\bigg [ \hat{\mathcal L}'_{{\mathbf{q}g}}  + \hat{\mathcal L}'^\dagger_{{\mathbf{q}g}}   \bigg], \\
 \hat{\mathcal L}^{({\rm o})}_{{\mathbf{q}g}} = \frac{i}{2} \bigg [\hat{\mathcal L}'_{{\mathbf{q}g}} - \hat{\mathcal L}'^{\dagger}_{{\mathbf{q}g}}   \bigg],
\end{split}
\end{align}
the two-body part of the Hamiltonian is expressed in a quadratic form
\begin{eqnarray}
\hat H_2' = \frac{1}{2} \sum_{\mathbf q}\sum_{g=1}^{n_{\mathbf q g}}  \bigg[ \hat {\mathcal L}^{(e)2}_{\mathbf{q}g}  + \hat {\mathcal L}^{(o)2}_{\mathbf{q}g} \bigg],
\end{eqnarray}
or in a more compact form
\begin{eqnarray}\label{E:H_quad_svd}
\hat H_2' = \frac{1}{2} \sum_{\mathbf q}\sum_{g=1}^{2n_{\mathbf q g}}  \hat {\mathcal L}^2_{\mathbf{q}g}.
\end{eqnarray}
Here, we also defined:
\begin{equation}\label{E:L_svd_fin}
    \hat{ \mathcal{L}}_{ \mathbf qg} = \begin{cases}
    \hat { \mathcal L}^{({\rm e})}_{{\mathbf qg}}   & \text{for}~{1 \le {g} \le n_{\mathbf {q }g}},\\
    \hat { \mathcal L}^{({\rm o})}_{{\mathbf qg'}}  & \text{for}~{n_{\mathbf {q }g}+1 \leq {g} \le { 2n_{\mathbf {q }g}}},
    \end{cases}
\end{equation}
where $g'=g-n_{\mathbf {q }g}$.

\subsection{Mean-field subtraction}\label{Se:mf_sub}
The $\hat {\mathcal L}_{\mathbf q g }$ operators induce density fluctuations.
Since the reference is the vacuum state, these fluctuations may be too strong and lead to large variances in the energy estimates or even hamper the convergence to the ground state.
Hence, shifting the reference to the mean field improves the numeric stability.
 The modified operators read  
\begin{align}\label{E:MF_shift}
\begin{split}
&\hat {\mathfrak L}_{ {\mathbf{q}g}} = \hat{\mathcal L}_{{\mathbf{q}g}} - \bar {\mathcal{ L}}_{g}\delta_{\mathbf q,0},
\end{split}
\end{align}
with 
\begin{eqnarray}\label{E:MF_expec}
\bar {\mathcal{L}}_{g} = \frac{\langle \Psi_{\rm T} | \hat {\mathcal{L}}_{{\mathbf{0}g}} | \Psi_{\rm T} \rangle}{\langle \Psi_{\rm T} | \Psi_{\rm T} \rangle},   
\end{eqnarray}
where the trial wavefunction $|\Psi_{\rm T}\rangle$ approximates the ground-state
wavefunction with a single Slater determinant.
Finally, the Hamiltonian is expressed as
\begin{eqnarray}\label{E:hamil_MF_fin}
\hat H = \hat H_1 + \hat H_2,
\end{eqnarray}
where
\begin{align}\label{E:H_MF_fin}
\begin{split}
\hat H_1 &= \hat H_1' +  \sum_{g} \bar {\mathcal L}_{g} 
\bigl( \hat {\mathcal L}_{0g} - \frac{1}{2} \bar{\mathcal{L}}_g\bigr),
\\
\hat H_2 &= \frac{1}{2} \sum_{\mathbf q} \sum_{g}  \hat{\mathfrak {L}}^{2}_{{\mathbf{q}g}}.
\end{split}
\end{align}

\subsection{Update procedure}\label{Sec:update}

Using a Hubbard-Stratonovich transformation,\cite{hubbard:1959} one can express the time-evolution for a single time step $\tau$ as (see Appendix \ref{Sec:B_appendix} for more details)
\begin{align}\label{E:one_body_propag}
\hat B(\mathbf x) =  \exp{\bigg(-\tau \hat H_1+i\sqrt{\tau} \sum_{\mathbf q} \sum_{g}x_{ \mathbf qg} \hat {\mathfrak L}_{ \mathbf qg}}\bigg),
\end{align}
where $\mathbf x$ is a random vector whose entries $x_{\mathbf qg}$ are normally distributed.
Thouless' theorem \cite{thouless:1960,thouless:1961} shows that $\hat B$ transforms a Slater determinant into another one.
Therefore, if one initializes an ensemble of $N_{\rm w}$ random walkers as single
Slater determinants $|\Psi_{\rm I}\rangle$, 
they remain single Slater determinants for all time steps $k$. 
The initial wavefunction $|\Psi_{\rm I}\rangle$ is set to the Hartee-Fock (HF) wavefunction\cite{hartree:1928,fock:1930,Slater:1930} denoted by $|\Psi_{\rm T}\rangle$.
In addition, we assign complex weights $W_{k}^w{\rm e}^{i\theta_{k}^w}$ to each of the walkers that are initialized to one. 
At the $k$-th MC step the walkers' states are updated as
\begin{eqnarray}\label{E:update_state}
|\Psi_{{k+1}}^w\rangle = \hat B(\mathbf x_{{k}}^w) |\Psi_{ {k}}^w\rangle.
\end{eqnarray}
The update criteria for the complex weights is 
\begin{eqnarray}\label{E:update_weight}
W_{{k+1}}^w{\rm e}^{i\theta_{{k+1}}^w} = \frac{\langle \Psi_{\rm T}| \Psi_{k+1}^w\rangle}{\langle \Psi_{\rm T}| \Psi_{k}^w\rangle} W_{k}^w{\rm e}^{i\theta_{k}^w}.
\end{eqnarray}
The update scheme formulated by Eqs.~\eqref{E:update_state} and \eqref{E:update_weight} is called free-projection AFQMC. This scheme is theoretically exact but numerically unstable due to the fermionic phase problem.\cite{Zhang:2003} The issue is suppressed by using the phaseless approximation.\cite{Zhang:2003} 
In this scheme, the walkers' states and weights are updated as\cite{motta_review:2018} 
\begin{align}\label{E:update_state_ph_afqmc}
\begin{split}
\centering
&|\Psi_{k+1}^w\rangle = \hat B(\mathbf x_{k}^w - {\mathbf f}_{k}^w) |\Psi_{ k}^w\rangle, \\
&W_{k+1}^w =  W_{k}^w \bigg |\frac{\langle \Psi_{\rm T} |  \Psi_{k+1}^w \rangle}{\langle \Psi_{\rm T} | \Psi_{k}^w \rangle} I_k^w \bigg|  \max(0,\cos(\Delta\theta)),
\end{split}
\end{align}
where $\Delta \theta$ is the phase difference of $\langle \Psi_{\rm T} |\Psi_{k+1}^w\rangle$ and $\langle \Psi_{\rm T}|\Psi_{k}^w\rangle$ at each time step, i.e.,
\begin{eqnarray}\label{E:delta_theta}
\Delta \theta = {\rm Arg} \bigg( \frac{\langle \Psi_{\rm T} |  \Psi_{k+1}^w \rangle}{\langle \Psi_{\rm T} | \Psi_{k}^w \rangle} \bigg),
\end{eqnarray}
and the importance sampling factor is defined as  
\begin{align}\label{E:import_func}
\begin{split}
I_k^w \equiv I(\mathbf x_{k}^w,  {\mathbf f_{ k}^w}, \Psi_{k}^w) 
= \exp\ \bigl[(\mathbf x_{k}^w - {\textstyle \frac12} \mathbf f_{k}^w) \cdot\mathbf f_{ k}^w\bigr].
\end{split}
\end{align}
This imposed shift $\mathbf f_{k}^w$ to the auxiliary field vector $\mathbf x_{k}^w$ is called force bias. Choosing the entries of the force bias vector as 
\begin{eqnarray}\label{E:force_bias}
f_{\mathbf qg,k}^w = -i \sqrt{\tau} \frac{\langle \Psi_{\rm T} | \hat{\mathfrak L}_{\mathbf qg} | \Psi_{k}^w \rangle}{\langle \Psi_{\rm T} | \Psi_{k}^w \rangle}
\end{eqnarray}
 minimizes the fluctuations in the importance function to first order in $\sqrt{\tau}$.\cite{motta_review:2018} See Appendix~\ref{Sec:fb_appendix} for the matrix representation of the force bias.

As the ph-AFQMC simulation proceeds the weights of some of the walkers become very large and statistically more important. 
The comb procedure\cite{Buonaura:1998,Booth_comb:2009} increases efficiency by splitting these walkers into multiple independent ones. Some walkers with small weight are killed to keep the total number of walkers constant.
The bias imposed by the population control can be removed by a standard linear extrapolation or using a large enough number of walkers in the simulation.\cite{Al-Saidi_dt_nw:2006,Purwanto_dt_nw:2009}

\subsection{Measurement of ground-state energy}
We measure the ground state energy at the $k$-th MC step in the ph-AFQMC simulation using
\begin{align}
E_{\rm 0} = \frac{\sum_{ kw} W_{ k}^w  E_{\rm{loc}}(\Psi_{k}^w)}{\sum_{ kw} W_{k}^w}~.
\end{align}
The local energy
\begin{align}\label{E:loc_energy}
E_{\rm{loc}}(\Psi_{k}^w) = \frac{\langle \Psi_{\rm T}| \hat H | \Psi_{k}^w\rangle}{\langle \Psi_{\rm T}| \Psi_{k}^w\rangle} = E_{1}(\Psi_{k}^w)+ E_{\rm H}(\Psi_{k}^w) + E_{\rm X}(\Psi_{k}^w)
\end{align}
consists of a single-particle contribution $E_1(\Psi_{k}^w)$ and the two-particle contributions of Hartree $E_{\rm H}(\Psi_{k}^w)$ and exchange $E_{\rm X}(\Psi_{k}^w)$.
We present the matrix representations of one- and two-body parts of the local energy in Appendix~\ref{Sec:loc_appendix}.

\subsection{Computational details}\label{Sec:comp_dets}
\begin{table}
\caption{\label{tab:pots} List of PAW potentials used in the present work
and the corresponding valence electrons.
For the projector radii $r_{\rm cut}$, the subscript describes their
angular momentum and a prefactor their multiplicity.
Within $r_{\rm core}$ the pseudopotential replaces the all-electron potential.
The values of $r_{\rm cut}$ and $r_{\rm core}$ are in atomic units.
In VASP, the PAWs are labeled Li\_GW\_sv for Li and \textit{X}\_GW for all other elements.
Potentials were released in the dataset potpaw\_PBE.54.}
\begin{ruledtabular}
\begin{tabular} {llll}
atom & valence & projector radii $r_{\rm cut}$ & $r_{\rm core}$ \\
\hline
Li & $1s^2 2s^1$ & $2 \times 1.2_s$, $1.3_s$, $2 \times 1.5_p$, $1.5_d$ & 1.5 \\
B  & $2s^2 2p^1$ & $2 \times 1.5_s$, $2 \times 1.7_p$, $1.7_d$ & 1.7 \\
C  & $2s^2 2p^2$ & $2 \times 1.2_s$, $2 \times 1.5_p$, $1.5_d$ & 1.5 \\
N  & $2s^2 2p^3$ & $2 \times 1.3_s$, $2 \times 1.5_p$, $1.5_d$ & 1.5 \\
F  & $2s^2 2p^5$ & $2 \times 1.1_s$, $2 \times 1.4_p$, $1.4_d$ & 1.4 \\
Ne & $2s^2 2p^6$ & $3 \times 1.4_s$, $3 \times 1.5_p$, $1.5_d$, $1.6_d$ & 1.6 \\
Al & $3s^2 3p^1$ & $2 \times 1.9_s$, $2 \times 1.9_p$, $2 \times 1.9_d$, $2.0_f$ & 2.0 \\
Si & $3s^2 3p^2$ & $2 \times 1.9_s$, $2 \times 1.9_p$, $2 \times 1.9_d$, $1.9_f$ & 1.9 \\
P  & $3s^2 3p^3$ & $2 \times 1.9_s$, $2 \times 1.9_p$, $2 \times 2.0_d$, $2.0_f$ & 2.0 \\
Ar & $3s^2 3p^6$ & $1.4_s$, $1.9_s$, $2 \times 1.9_p$, $2 \times 1.9_d$, $1.9_f$ & 1.9 \\
\end{tabular}
\end{ruledtabular}
\end{table}
\newcolumntype{.}[1]{D{.}{.}{#1}}
\begin{table}
\caption{\label{tab:lat_const} Experimentally measured or extrapolated $T=0$~K lattice constants $a$, crystal structures and plane-wave energy cutoffs $E_{\rm cut}$ for the solids studied in the present work. All plane-wave energy cutoffs are in eV and the lattice constants are in \AA.}
\begin{ruledtabular}
\begin{tabular}{cccc}
crystal & symmetry & $a$ & $E_{\rm cut}$\\ \hline
Ne  & Fm$\bar3$m  &  4.430  \cite{wyckoff1963crystal} & 1000 \\ 
Ar  & Fm$\bar3$m  &  5.260 \cite{wyckoff1963crystal} & 1000\\  
C   & Fd$\bar3$m  &  3.567 \cite{Staroverov:2004}  & 1000  \\
SiC & F$\bar4$3m  &   4.358 \cite{Staroverov:2004} & 1000\\  
Si   & Fd$\bar3$m  &  5.430 \cite{Staroverov:2004}  & 1000\\ 
LiF & Fm$\bar3$m  &   4.010 \cite{Staroverov:2004} & 1500\\ 
LiCl & Fm$\bar3$m  &  5.106 \cite{Staroverov:2004}  & 1500\\  
BN  & F$\bar4$3m  &   3.607 \cite{madelung2004semiconductors}  & 1000\\  
BP  & F$\bar4$3m  &   4.538 \cite{madelung2004semiconductors} & 1000 \\ 
AlN  & F$\bar4$3m  &  4.380 \cite{trampert1997crystal} & 1000 \\ 
AlP  & F$\bar4$3m  &  5.460 \cite{madelung2004semiconductors}  & 1000 \\ 
\end{tabular}
\end{ruledtabular}
\end{table}
We used the Vienna Ab initio Simulation Package\cite{Bloechl1994,Kresse1999} (VASP) 
to compute the matrix representation of $\hat H'_1$ in Eq.~(\ref{E:H_1}) and $\hat L'_{\mathbf{qG}}$ in Eq.~(\ref{E:H_2}).
VASP represents the pseudo-orbitals with plane waves but computes the matrix elements
with all-electron precision using the PAW method.
For more details about the VASP interface we refer reader to Appendix~\ref{appendix:VASP}.
We employed the PAW potentials listed in Table.~\ref{tab:pots}.
For Li, we considered the semi-core $1s$ states as valence states. 
Table~\ref{tab:lat_const} summarizes the experimental lattice constants, the crystal structures and the plane-wave energy cutoffs of the solids explored in Sec.~\ref{Sec:result}.
We computed the MP2, CCSD, and CCSD(T) energies with the same setup in VASP to ensure comparability.

The probe-charge Ewald method treats the Coulomb kernel
singularity in the reciprocal space.\cite{Massidda:1993} 
In this method, one subtracts an auxiliary function, which has the same singularities as the Coulomb kernel, to regularize the exchange integral. \cite{Gygi:1986} 
In practice, VASP calculates the correction by placing a probe-charge and compensating homogeneous background into a supercell (determined by the primitive cell and the employed $\mathbf k$-point grid) and calculates this energy using an Ewald summation.\cite{Paier:2005} 
This introduces a constant shift in the input matrices, which we call singularity correction. 
The singularity correction must be taken into account in the calculation of total energies but does not affect energy differences. For instance, the
HF energy and the ph-AFQMC energies shift by a constant value proportional to the number of electrons and the singularity correction. This means that we do not need to apply the correction in the ph-AFQMC calculation, as long as we determine only the {\em correlation} energy in the ph-AFQMC calculation. We also performed ph-AFQMC calculations with and without imposing the singularity correction and confirmed that both approaches yield identical correlation energies within statistical errors.

Furthermore, as a sanity test, we reconstructed the Fock matrix from the input matrices and ascertained that they yield consistent eigenvalues and MP2 correlation energies. 

\section{Results}\label{Sec:result}
In this section, we present our ph-AFQMC results.
We explore the errors associated with the finite time step and the population control (Sec.~\ref{Sec:errors}).
Comparing the correlation energy, we validate the implementation with a $\mathbf k$-point
mesh and the equivalent supercell (Sec.~\ref{Sec:supercell}).
For the equation of state of diamond, ph-AFQMC agrees almost perfectly with CCSD(T) much improving on CCSD and MP2 (Sec.~\ref{Sec:EoS}).
This trend holds for the correlation energy of many more materials (Sec.~\ref{Sec:corr_en}).
Finally, we employ a down-sampling technique in order to recover the correlation energy of a denser $\mathbf k$-point mesh at large basis sets (Sec.~\ref{Sec:downsampling}). All energies are reported in eV per unit cell.

\subsection{Population-control and time-step errors}\label{Sec:errors}

\begin{figure}
\centering
\includegraphics[width=1.0\linewidth]{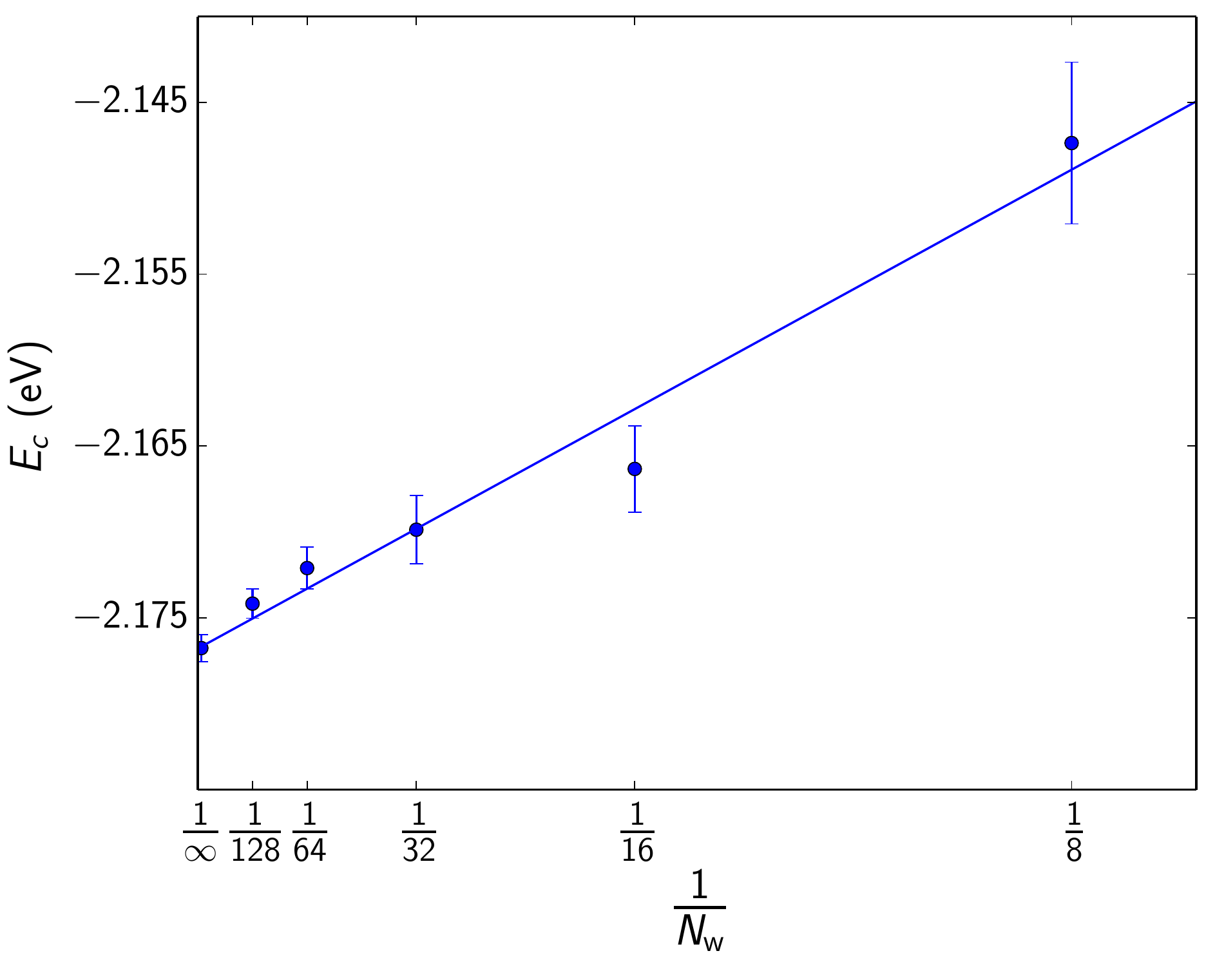}
\caption{ph-AFQMC correlation energy $E_{\rm c}$ vs. inverse population size for diamond using a $\Gamma$-centered $2\times2\times2$ 
$\mathbf k$-point mesh with 8 electrons in 8 HF orbitals per $\mathbf k$ point. 
The population control biases $E_{\rm c}$ when small populations are used.
256 walkers reproduce $E_{\rm c}$ of the infinite population size limit to a precision of 1~meV.
\label{fig:C_222_WALKERS}}
\end{figure}
\begin{figure}
\centering
\includegraphics[width=1.0\linewidth]{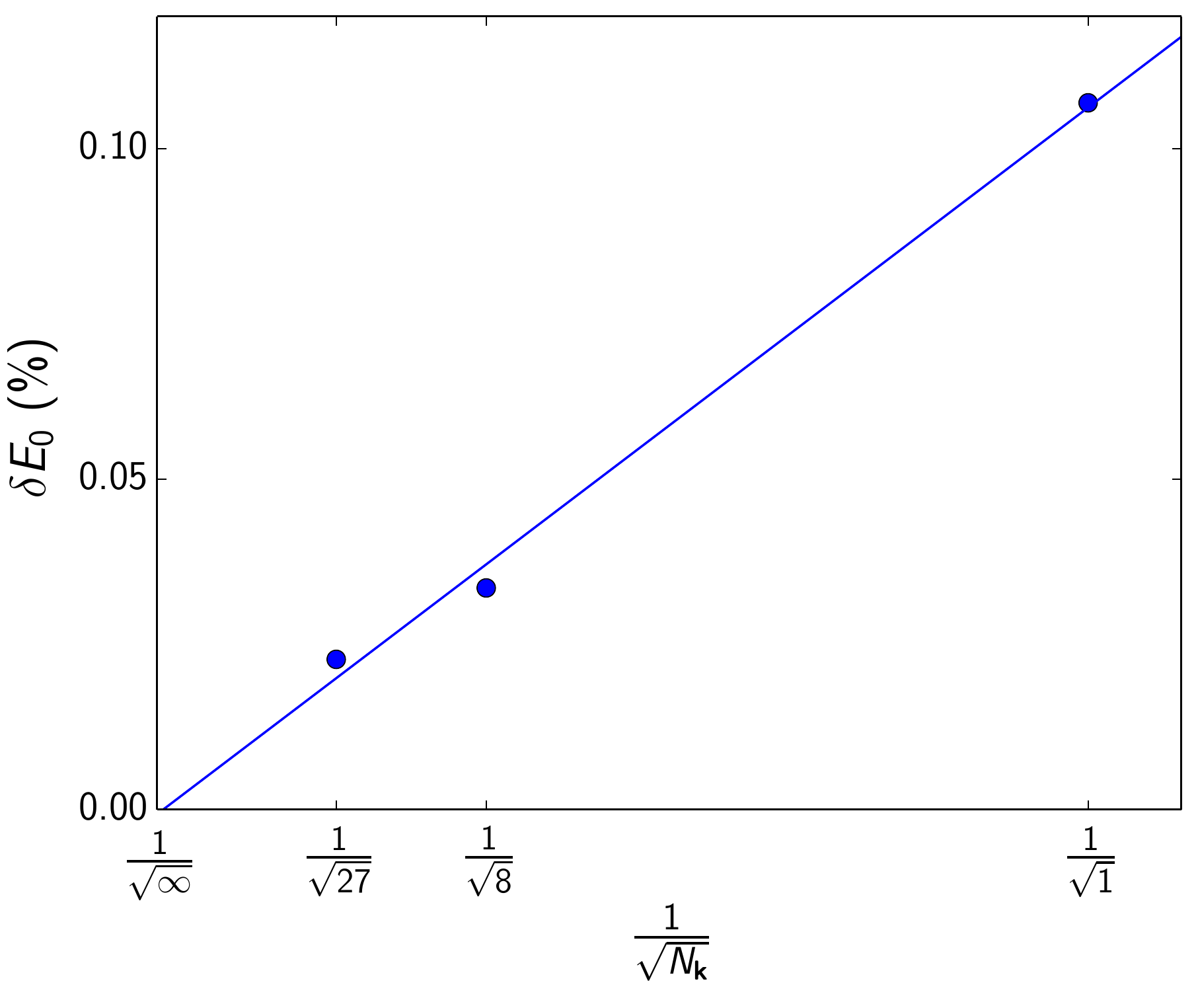}
\caption{In diamond, the relative error $\delta E_0$ of the total 
energy shows a linear relation with $\frac{1}{\sqrt{N_{\mathbf k}}}$ where $N_{\mathbf k}$ is the number of $\mathbf k$ points. We employed 8 HF orbitals per $\mathbf k$ point. }
\label{fig:rel_err}
\end{figure}
\begin{figure}
\centering
\includegraphics[width=1.0\linewidth]{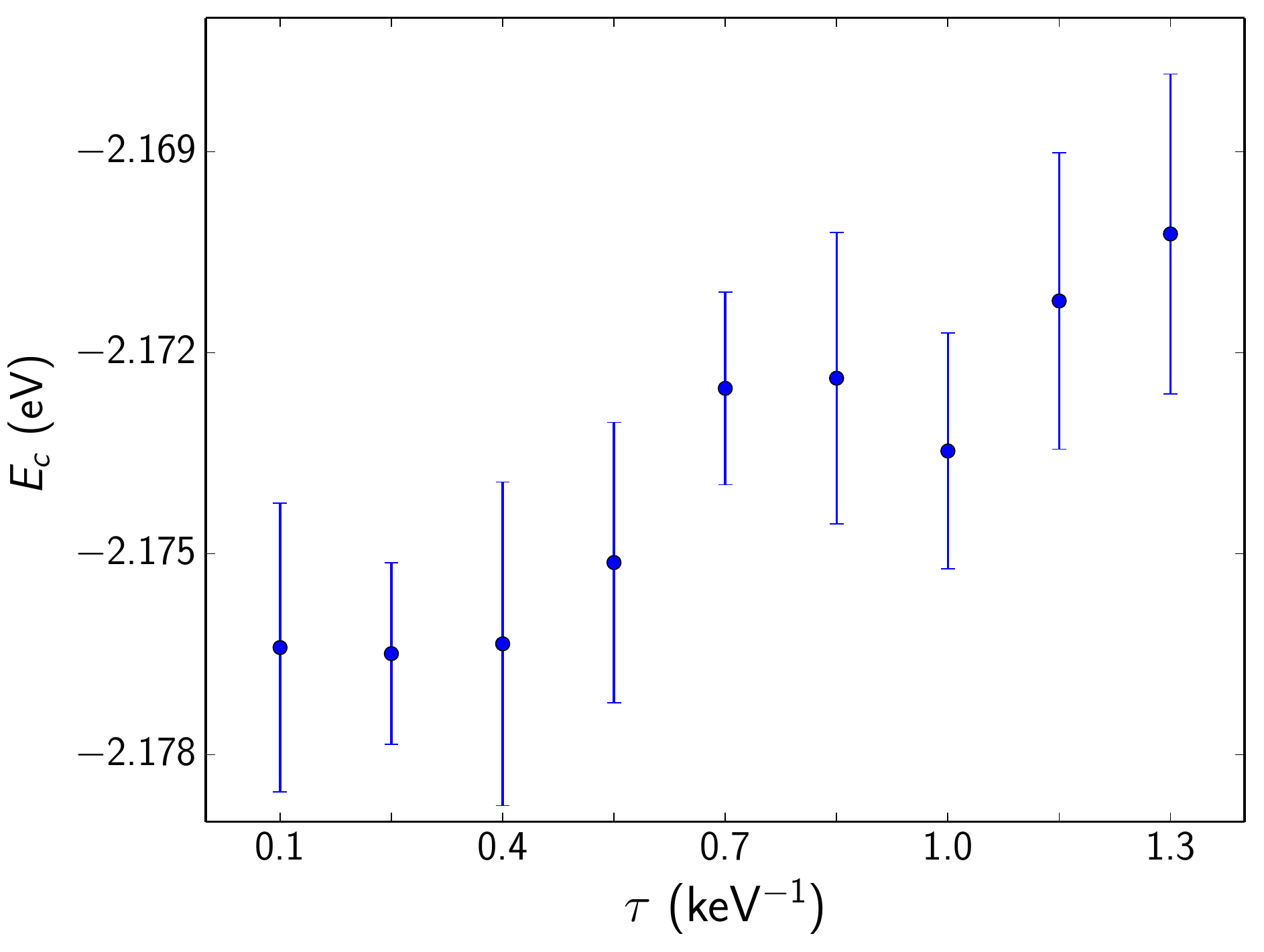}
\caption{ph-AFQMC correlation energy $E_{\rm c}$ vs. time step $\tau$ for diamond using a $\Gamma$-centered
$2\times2\times2$ $\mathbf k$-point mesh with 8 electrons in 8 HF orbitals per $\mathbf k$ point.
For all $\tau \leq 0.55$~${\rm keV}^{-1}$, the calculated $E_{\rm c}$ agree within statistical errors.
\label{fig:C_222_TAU}}
\end{figure}
We investigate the time-step error and the population-control bias to determine appropriate choices for the time step and the walker population size to ascertain that these systematic errors are smaller than the statistical error.

For small walker populations, the population control biases the correlation energy towards smaller absolute values.\cite{motta_review:2018}
Fig.~\ref{fig:C_222_WALKERS} shows the correlation energy $E_{\rm c}$ of diamond employing a $2\times2\times2$ $\mathbf k$-point mesh.
The error in the correlation energy increases proportionally to the inverse of the population size
consistent with what Ref.~\onlinecite{motta_review:2018} reports for molecules.
This dependence allows to extrapolate to the infinite population size to correct for this bias.\cite{Al-Saidi_dt_nw:2006,Purwanto_dt_nw:2009}
The linear extrapolated value agrees with the result with 2048 walkers. 
We find that 256 walkers are sufficient to calculate the correlation energy to a precision of 1~meV. 
 
Fig.~\ref{fig:rel_err} shows how the relative error 
$\delta E_0$ of the total energy per unit cell scales with number of $\mathbf k$
points $N_{\mathbf k}$. 
Here, we defined $\delta E_0 = \frac{\Delta E_0}{E_0}$, where $\Delta E_0$ is the standard error of the mean.
We conclude that utilizing a denser $\mathbf k$-point mesh requires a smaller number of MC steps because each walker samples more of the Brillouin zone in each MC step.
For a fixed number of walkers, we observe a linear relationship between the relative error and $\frac{1}{\sqrt{N_{\mathbf k}}}$.
Therefore, denser $\mathbf k$-point meshes achieve the same relative
error $\delta E_0$ in $\frac{1}{\sqrt{N_{\mathbf k}}}$ fewer MC steps, if the number of walkers is kept fixed. This observation is very beneficial,
and suggests that the statistical errors at different $\mathbf k$ points are largely uncorrelated and thus average out. It also means that we can either  decrease the number of walkers or the sampling period by $N_{\mathbf k}$, when the number of $\mathbf k$ points increases.

ph-AFQMC is only accurate up to linear order in the time step $\tau$.
The Hubbard-Stratonovich transformation, the Trotter factorization\cite{trotter:1959}, and evaluating the matrix exponential introduce time-step errors.\cite{motta_review:2018}
Fig.~\ref{fig:C_222_TAU} illustrates that time steps $\tau \leq 0.55$~${\rm keV}^{-1}$ yield equivalent correlation 
energies considering the statistical fluctuations.
Larger time steps systematically decrease the absolute value of the correlation energy.
In this work, we set the time step to $\tau = 0.25 $~keV$^{-1}$ and expect a negligible (less than statistical error) 
time-step error without extrapolation.

\subsection{Supercell vs. primitive-cell calculation} \label{Sec:supercell}

\newcolumntype{.}[1]{D{.}{.}{#1}}
\begin{table}
\caption{\label{tab:prim_sc}ph-AFQMC correlation energies of the primitive cell 
($E^{\rm AF}_{\rm prim}$) and of the corresponding supercell ($E^{\rm AF}_{\rm sc}$) agree within statistical fluctuations.
The absolute value of ph-AFQMC correlation-energy is systematically larger than the one obtained with 
CCSD(T) ($E_{\rm sc}^{\rm CC}$). }
\begin{ruledtabular}
\begin{tabular}{l l .{3.9} .{3.8} .{2.4} }
\multicolumn{2}{c}{crystal} & \multicolumn{1}{c}{$E^{\rm AF}_{\rm prim}$(eV)} & \multicolumn{1}{c}{$E^{\rm AF}_{\rm sc}$(eV)} & \multicolumn{1}{c}{$E^{\rm CC}_{\rm sc}$(eV)} \\ \hline
Ne  & Fm$\bar3$m  &  -0.2131(12)   &  -0.2127(12)  &  -0.2107  \\
C   & Fd$\bar3$m  &  -1.4945(101)  &  -1.4951(54)  &  -1.4830  \\
BN  & F$\bar4$3m  &  -1.1913(153)  &  -1.1988(52)  &  -1.1861  \\
LiF & Fm$\bar3$m  &  -0.5257(18)   &  -0.5223(37)  &  -0.5134  \\
SiC & F$\bar4$3m  &  -1.0323(76)   &  -1.0310(70)  &  -1.0252  \\
\end{tabular}
\end{ruledtabular}
\end{table}
In this subsection, we validate our ph-AFQMC implementation with a $\mathbf k$-point mesh in the primitive cell against the corresponding supercell. 
Sampling the Brillouin zone with a $2\times2\times2$ $\mathbf{k}$-point mesh is 
equivalent to a $\Gamma$-point calculation of a supercell enlarged by a factor of two along
each axis. The primitive-cell calculation takes advantage of translational symmetry.
Therefore, the number of numerical operations is roughly a factor $N_{\mathbf k}$ smaller in the primitive cell than in the corresponding supercell. We note in passing that these savings are not always observed in actual calculations, since the matrix dimensions are significantly smaller in the primitive cell, resulting in some performance loss.

For a direct comparison, it is important that the primitive cell and the supercell use the same orbitals.
The orbitals are ordered by their eigenvalues $\varepsilon_{n\mathbf k}$
at each $\mathbf k$ point.
In the primitive cell, $\varepsilon_{n+1\mathbf k} < \varepsilon_{n \mathbf k'}$ is possible; whereas in the supercell the former would be included before the latter.
To address this, we include more bands in the primitive cell such that all eigenvalues of the supercell are reproduced.
Then, we zero all contributions in the primitive cell corresponding to orbitals not present in the supercell.

In Table~\ref{tab:prim_sc}, we verify that our ph-AFQMC implementation computes correlation
energies consistent within statistical fluctuations for five different crystals.
For all systems, we used 8 HF orbitals per $\mathbf k$ point in the primitive cell.
Compared to ph-AFQMC, CCSD(T) yields systematically about 10~meV more positive correlation energies, 
a point we will return to in Section \ref{Sec:corr_en}.   
\subsection{Equation of state} \label{Sec:EoS}
\begin{figure}
\centering
\includegraphics[width=1.0\linewidth]{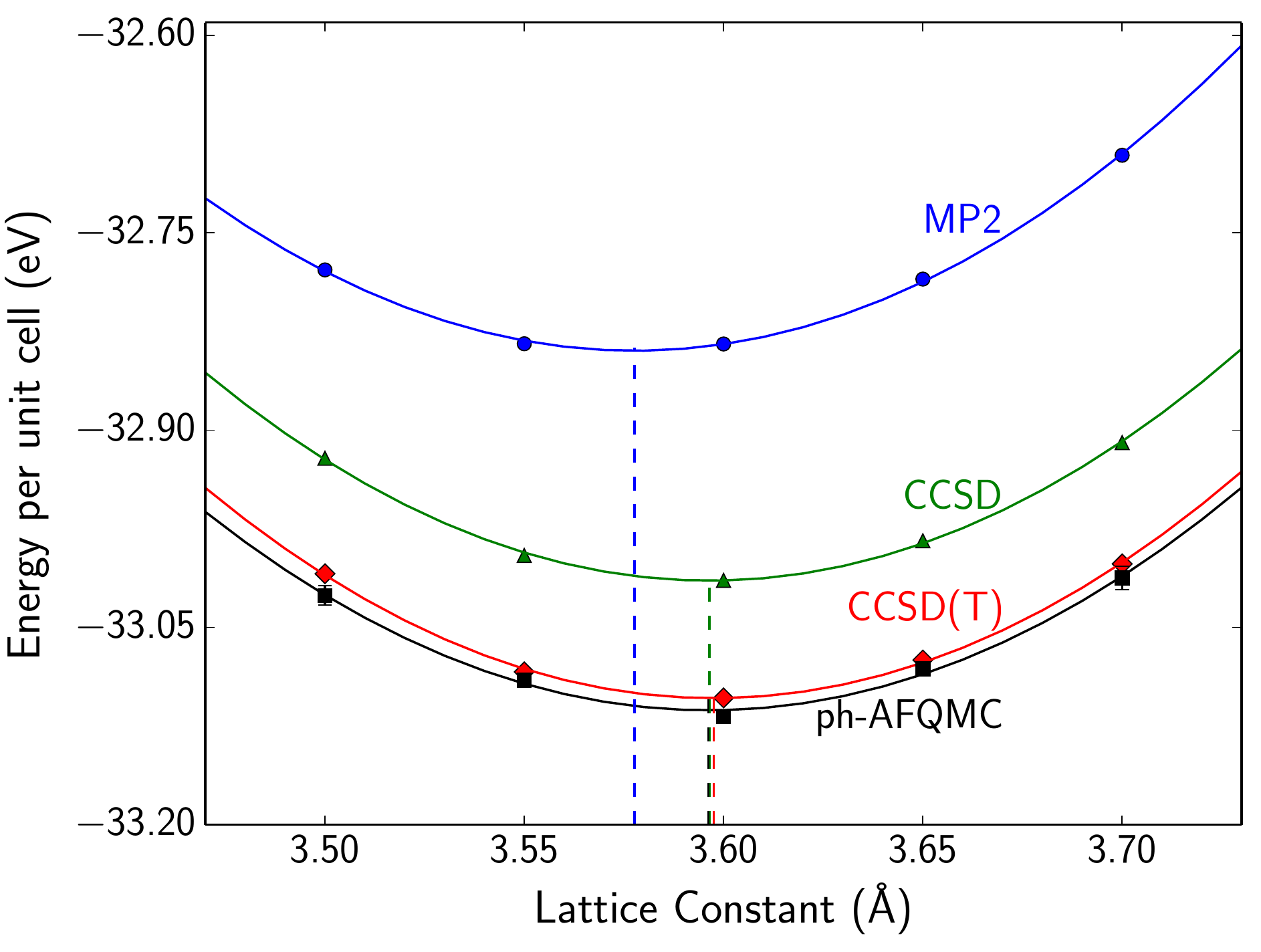}
\caption{
Equation of state for diamond: MP2 (blue) and CCSD (green) yield larger energies than
CCSD(T) (red) and ph-AFQMC (black). The optimized lattice constants (dashed vertical lines) are very similar except for the MP2 method. CCSD and ph-AFQMC vertical lines, marking the equilibrium lattice constants, are visually indistinguishable.
\label{fig:Etot_mp2_afqmc_ccsd_C_vs_a}}
\end{figure}
In this section, we benchmark the accuracy of the ph-AFQMC total energies against popular deterministic quantum-chemistry methods for diamond, concentrating in particular on relative energies, such as those produced by changes of the volume. 
Fig.~\ref{fig:Etot_mp2_afqmc_ccsd_C_vs_a} shows the equation of state of diamond for a range of lattice constants.
The calculations use a $\Gamma$-centered $3\times3\times3$ $\mathbf k$-point mesh with 8 HF orbitals per $\mathbf k$ point.
Compared to ph-AFQMC, the total energies of MP2 and CCSD are 273~meV and 99~meV higher, respectively.
In contrast, the CCSD(T) energies are less than 10~meV higher than the ph-AFQMC energies. 
Furthermore, the optimized lattice constant is 3.578~\AA~in MP2 considerably
smaller than the one of 3.596~\AA~in ph-AFQMC.
CCSD and CCSD(T) yield lattice constants of 3.596~\AA~and 3.597~\AA, respectively,
almost identical to the ph-AFQMC result.

\subsection{Comparison with coupled-cluster methods for total energies}\label{Sec:corr_en}
\begin{figure}
\centering
\includegraphics[width=1.0\linewidth]{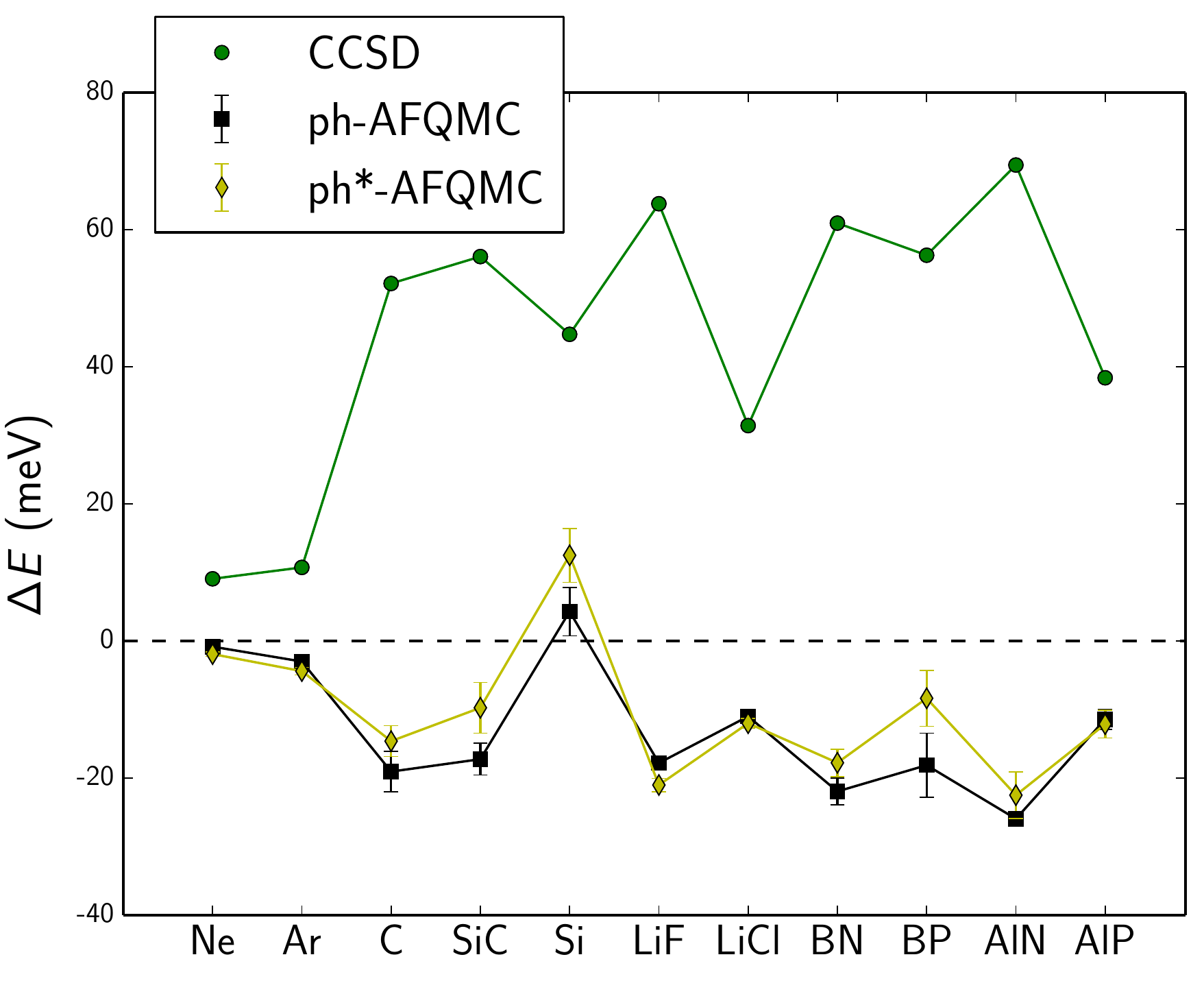}
\caption{Relative correlation energies obtained from CCSD (green), ph-AFQMC (black) and ph*-AFQMC (yellow) with respect to the CCSD(T) values for a range of crystals employing
a $\Gamma$-centered $2\times2\times2$ $\mathbf k$-point mesh. The reference values are the CCSD(T) correlation energies, i.e., $\Delta E = E_{\rm c} - E_{\rm c}^{\rm CCSD(T)}$.
\label{fig:re_abs_std_cc_ev}}
\end{figure}
This subsection illustrates what accuracy of the correlation energy one can expect from ph-AFQMC compared to coupled-cluster methods.
We compute correlation energies for several prototypical semiconductors and insulators with experimental band gaps ranging
from 1.4~eV (Si) to 21.7~eV (Ne) to cover different bonding situations.
A $\Gamma$-centered $2\times2\times2$ $\mathbf k$-point mesh allows for a reasonable run time for all systems.
For LiF and LiCl, we use 10 electrons and 9 HF orbitals per $\mathbf k$ point.
All other materials are calculated with 8 valence electrons and 8 HF orbitals per $\mathbf k$ point. 

We compare the correlation energy of the different methods in Fig.~\ref{fig:re_abs_std_cc_ev}.
With the exception of the noble-gas crystals, CCSD underestimates the absolute value of the correlation energy by
40~meV to 95~meV compared to ph-AFQMC.
In contrast, ph-AFQMC and CCSD(T) agree within 25~meV for the correlation energy of all materials, which
is within chemical accuracy.

Sukurma \emph{et al.}\cite{Zoran:2023} recently proposed a modified version of the phaseless approximation (ph*-AFQMC) to suppress the over-correlation issues of ph-AFQMC. 
In this approach, the complex nature of the walker weights is retained and if $|\theta_k^w| \ge \frac{\pi}{2}$ the walker is explicitly killed. Furthermore, the real part of the importance weight instead of the absolute value is used in the update procedure. 
For more detailed description of ph*-AFQMC we refer the reader to Ref.~\onlinecite{Zoran:2023}. 
Fig.~\ref{fig:re_abs_std_cc_ev} demonstrates that ph*-AFQMC yields either consistent or slightly less negative correlation energies compared to the original ph-AFQMC  method for the explored solids with the exception of LiF. 
This is consistent with what Ref.~\onlinecite{Zoran:2023} reports for the HEAT set molecules.       

Generally, compared to CCSD(T), ph-AFQMC yields more negative correlation energies with the exception of silicon.
As to why the ph-AFQMC values are more negative than the CCSD(T) values, we need to speculate.
Generally, for weakly correlated systems, coupled-cluster methods converge from above to the exact correlation
energy.\cite{Bomble:2005} 
In other words, CCSD(T) has a slight tendency to under-correlate compared to coupled-cluster methods that include triple, quadruple and pentuple excitation operators.\cite{Bomble:2005} 
However, likewise ph-AFQMC can also yield too negative correlation energies.\cite{Zoran:2023} 
Only if strong double excitations are relevant, ph-AFQMC generally tends to under-correlate.
Since both methods are non-variational, it is hard to tell which value is more accurate.  
The under-correlation for Si and the relatively small single particle band gap of Si, however,
suggests that the slight under-correlation for Si is related to ph-AFQMC underestimating
energy contributions from double excitations.

Since the ph*-AFQMC correlation energies are also more negative than the CCSD(T) ones, we tend to believe that the ph*-AFQMC are closer to the ground truth. 
Anyhow, to resolve this issue, obviously more accurate reference methods are required.
Booth \emph{et al.}\cite{booth:2013} employed FCIQMC with the initiator approximation
($i$-FCIQMC) on the same set of materials to investigate the accuracy of standard quantum-chemistry methods. 
However, recent work suggests that various approximations such as the number of initiators used in this seminal work could have led to an underestimation of the correlation energies.\cite{Ghanem:2019}
Underestimated absolute correlation energies were also found for benzene, where $i$-FCIQMC results were 1.5 kcal/mol above the most accurate total energy estimates.\cite{eriksen2020ground,Blunt:2019}
Hence, although our results are within chemical accuracy of the reference values, fully converged CI calculations with consistent basis sets and pseudopotentials are required to resolve this ambiguity.
In this context it is also worth noting that Lee \emph{et al.}\cite{Lee:2019}
carried out a comparative study using ph-AFQMC for the uniform electron gas.
Their findings show that ph-AFQMC correlation energies are in good agreement
with  $i$-FCIQMC at densities corresponding to $r_S=1.0$ and $r_S=2.0$,
whereas at $r_S=5.0$ the $i$-FCIQMC correlation are more negative than those obtained
with ph-AFQMC.\cite{Lee:2019,Shepherd:2012}
CCSDT theory yields energies in good agreement with $i$-FCIQMC at high densities but
underestimates the absolute correlation energies at densities corresponding $r_S=2.0$ and  $r_S=5.0$.\cite{Lee:2019,Neufeld:2017} 
Although CCSD(T) and CCSDT can differ, this is not expected
for the investigated relatively small system sizes with about 14 electrons.
Therefore, the uniform electron gas findings also support our hypothesis that ph*-AFQMC is closer
to the ground truth for the investigated solids in this work.

\begin{figure}
\centering
\includegraphics[width=1.0\linewidth]{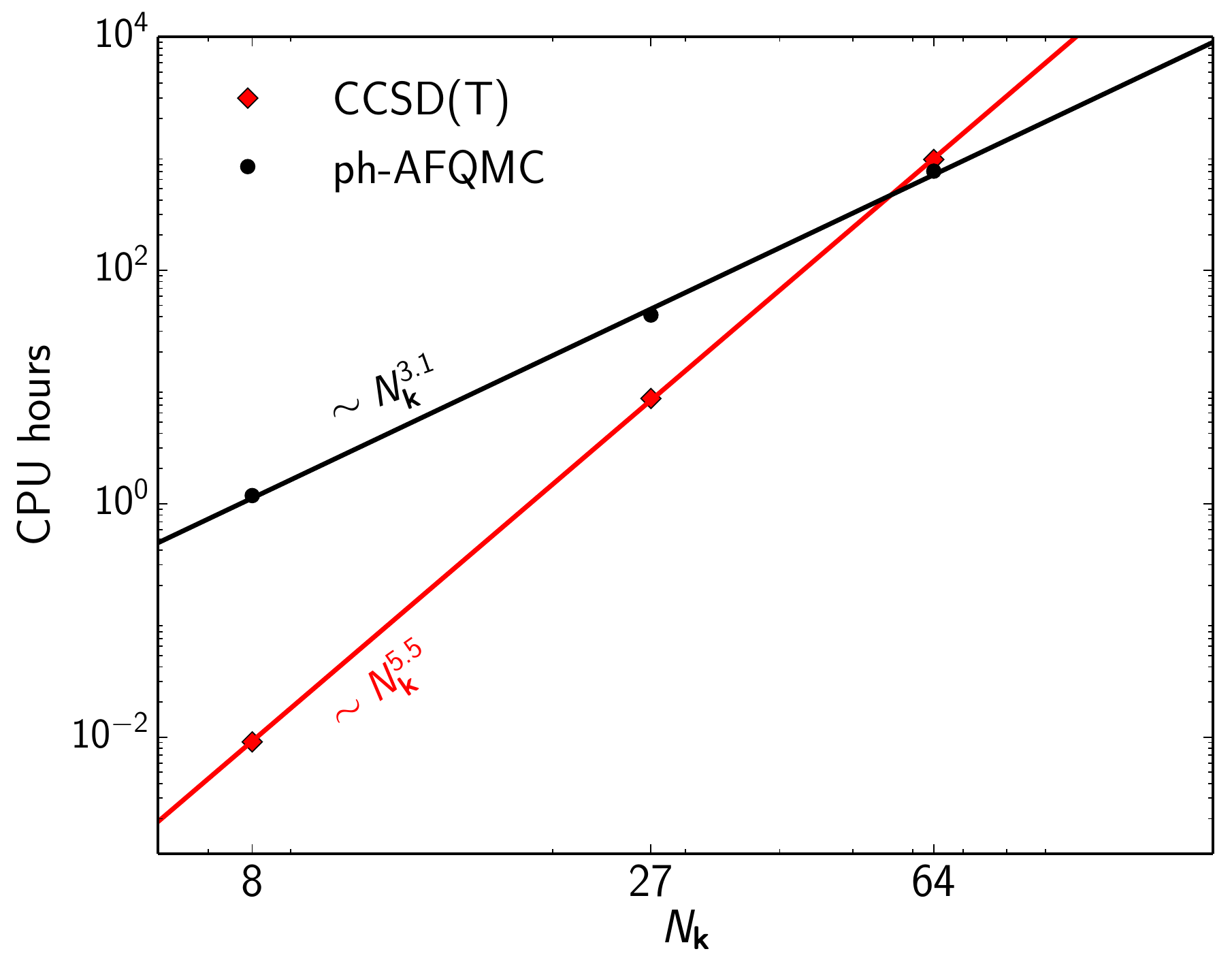}
\caption{System size scaling of ph-AFQMC (black) and CCSD(T) (red) for various $\Gamma$-centered $\mathbf k$-point meshes of diamond. We employed 8 HF orbitals per $\mathbf k$ point. Both ph-AFQMC and CCSD(T) calculations were performed on a single-socket AMD EPYC 7713 and one MPI process.}
\label{fig:scaling}
\end{figure}
ph-AFQMC exhibits a better scaling with system size compared to CCSD(T).
To illustrate this, we benchmark the scaling of our ph-AFQMC Python code against the VASP CCSD(T) code developed in Fortran.
In Fig.~\ref{fig:scaling}, we explore the scaling for various regular 
$\Gamma$-centered $\mathbf k$-point meshes of diamond with 8 orbitals per $\mathbf k$ point. 
The CCSD(T) computational cost is slightly larger than for ph-AFQMC at $N_{\mathbf k}=64$, which is the largest system size we have investigated in this work. 
Empirically, ph-AFQMC shows a roughly-cubic scaling with $N_{\mathbf k}$ while CCSD(T) approximately scales as $N_{\mathbf k}^{5.5}$, here.

\subsection{Down-sampling of the correlation energy}\label{Sec:downsampling}
\begin{figure}
\centering
\includegraphics[width=1.0\linewidth]{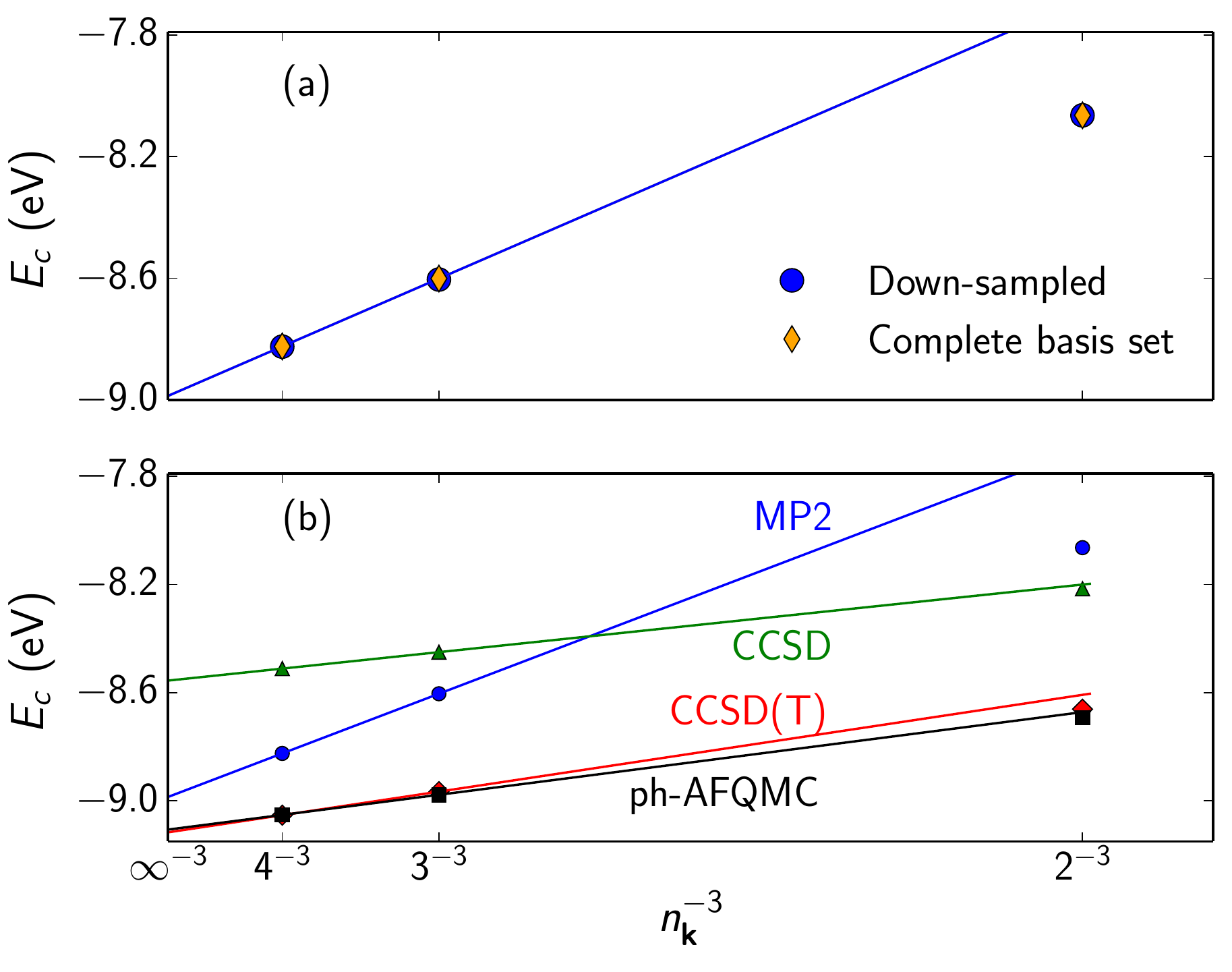}
\caption{
Diamond correlation energy $E_{\rm c}$ sampled with an $n_{\mathbf k}\times n_{\mathbf k}\times n_{\mathbf k}$
$\mathbf k$-point mesh.
The solid lines extrapolate to $n_{\mathbf k} \rightarrow \infty$ using $n_{\mathbf k} = 3$ and $n_{\mathbf k} = 4$.
(a) Comparison of the down-sampled correlation energy ({blue}) with the one obtained with the complete basis set for MP2
({orange}). {The extrapolated lines are visually indistinguishable.} (b) The down-sampled correlation energies for MP2 (blue), CCSD (green), CCSD(T) (red) and ph-AFQMC (black).}
\label{fig:downsampling_diamond}
\end{figure}
The preceding subsections illustrated benchmark calculations for the PAW method with coarse $\mathbf k$-point meshes and
few HF orbitals.
In this subsection, we present a strategy
to perform predictive ph-AFQMC calculations. This strategy is outlined and tested for diamond. We test two commonly used techniques to converge to the complete basis-set (CBS) limit.
First, natural orbitals \cite{Lowdin:1955,Bender:1966,DAVIDSON:1972,Thunemann:1977} reduce the number of virtual orbitals required
to converge to the correlation energy.
We use natural orbitals obtained by the random phase approximation.\cite{Ramberger:2019,Moritz:2022}
Second, because differences converge faster than absolute energies, \emph{down-sampling} via energy differences\cite{Gruneis_NO:2011} 
allows to estimate the CBS correlation energy at significantly reduced cost.

Specifically, we intend to extrapolate the correlation energy to infinitely-dense $\mathbf k$-point meshes.
Here, we use $n_{\mathbf k} \times n_{\mathbf k} \times n_{\mathbf k}$ $\mathbf k$-point meshes for 
a linear fit to zero volume per $\mathbf k$ point based on $n_{\mathbf k} =3$ and $n_{\mathbf k} = 4$.
To obtain the down-sampled correlation energy at a denser $\mathbf k$-point mesh, we compute the energy difference when
increasing $n_{\mathbf k}$ at fixed number of orbitals per $\mathbf k$ point ${n}_{\rm b}$
\begin{equation}
    \Delta_{\rm c}(n_{\mathbf k}, {n}_{\rm b}) = 
    E_{\rm c}(n_{\mathbf k} + 1, {n}_{\rm b}) - E_{\rm c}(n_{\mathbf k}, {n}_{\rm b})~.
\end{equation}
We converge the correlation energy at the coarser $\mathbf k$-point mesh with respect to the number of orbitals per $\mathbf k$ point ${n}_{\rm b}$. 
Adding the difference yields an approximation for the correlation energy at the dense mesh
\begin{equation}
    E_{\rm c}(n_{\mathbf k} + 1) \approx \tilde{E}_{\rm c}(n_{\mathbf k} + 1,{n}_{\rm b}) = E_{\rm c}(n_{\mathbf k}) + \Delta_{\rm c}(n_{\mathbf k}, {n}_{\rm b}).
\end{equation}
One can then iterate this procedure with a reduced ${n}_{\rm b}$ to obtain approximations for even denser meshes.

To verify this procedure, we evaluate the correlation energy of MP2 with the down-sampling technique
\begin{align}
\begin{split}
    E_{\rm c}(n_{\mathbf k}=3)
    &
    \approx E_{\rm c}(n_{\mathbf k}=2) + \Delta_{\rm c}(n_{\mathbf k}=2, {n}_{\rm b}=16),
    \\
    E_{\rm c}(n_{\mathbf k}=4)
    &
    \approx E_{\rm c}(n_{\mathbf k}=3) + \Delta_{\rm c}(n_{\mathbf k}=3, {n}_{\rm b}=8)~.
\end{split}
\end{align}
We compute the correlation energy $E_{\rm c}(n_{\mathbf k} = 2)$ with 64 natural orbitals.
Fig.~\ref{fig:downsampling_diamond}a compares these energies to the CBS where all calculations use ${n}_{\rm b} = 64$;
the two sets of data points are visually indistinguishable.
Extrapolating to {the} infinitely dense $\mathbf k$-point mesh yields a correlation energy of $-$8.987~eV for the CBS and of
$-$8.986~eV for the down-sampled energies.

We employ the down-sampling technique for the correlation energy of MP2, CCSD, CCSD(T), and ph-AFQMC.
Fig.~\ref{fig:downsampling_diamond}b shows the resulting extrapolations.
For MP2, the extrapolation from $n_{\mathbf k} = 3$ to $n_{\mathbf k} = 4$ shows a different slope and
a large deviation for $n_{\mathbf k} = 2$ compared to the other methods.
In contrast, the ph-AFQMC correlation energy would only change by 9~meV if one used $n_{\mathbf k} = 2$
and $n_{\mathbf k} = 3$ for the extrapolation.
CCSD(T) and ph-AFQMC agree within statistical errors for denser meshes. 
The extrapolated correlation energy is $-$9.116~eV for CCSD(T) and $-$9.106(30)~eV for ph-AFQMC.

\begin{table}
\caption{\label{tab:diamond}CCSD(T) and ph-AFQMC down-sampled correlation energies of the primitive cell
for various $n_{\mathbf k}$ and $n_{\rm b}$.}
\begin{ruledtabular}
\begin{tabular}{l r r r  }
\multicolumn{1}{c}{$n_{\mathbf k}$} & \multicolumn{1}{c}{$n_{\rm b}$} & \multicolumn{1}{c}{ $E_{\rm c}^{\rm CC}$~(eV)} & \multicolumn{1}{c}{ $E_{\rm c}^{\rm AF}$~(eV)}  \\ \hline
2  &  64   & -8.661   & -8.693(10)\\ 
3  &  16   & -8.966   & -8.977(13)\\
4  &   8   & -9.052   & -9.052(16)\\
\end{tabular}
\end{ruledtabular}
\end{table}

There are two important conclusions we can draw from this test, and these are also more clearly borne out in Table  \ref{tab:diamond}. First, the difference in the correlation energy between ph-AFQMC and CCSD(T)
does not increase as the number of virtual orbitals increases. 
On the contrast it seems to decrease, although we might need better statistical accuracy and tests for more materials to be certain.
This observation is fully in line with the  observations we recently made for small molecules.\cite{Zoran:2023} 
This also means that the potential errors of both methods relate to low energy excitations, and tests using few states are already very meaningful. 
Second, increasing the number of ${\mathbf k}$ points does not change the difference between ph-AFQMC and CCSD(T) appreciably (note that our error bars for the densest ${\mathbf k}$-point mesh are fairly sizable).
Hence, the excellent agreement of CCSD(T) and ph-AFQMC prevails even in the thermodynamic limit. 

\section{Conclusion}\label{Sec:conclusion}

ph-AFQMC is potentially a great method to obtain very accurate reference results for solid-state systems. 
The present work merely tries to establish that ph-AFQMC is competitive to CCSD(T) and is capable to yield very accurate results for solids with quite different characteristics. 
As already emphasized in the introduction, one advantage of ph-AFQMC is that it is fully compatible with other quantum-chemistry methods, such as MP2, CCSD or CCSD(T). 
This means that one can directly compare predicted energies with these and other well-established quantum-chemistry methods. 
It is clear and well understood that this also entails significant disadvantages, such as a slow convergence with respect to the number of virtual states (unoccupied orbitals) included in the calculations. 
Certainly DMC is superior in this respect, but validation of DMC against other quantum-chemistry methods is notoriously difficult. 
For instance, recent conflicting results for localized coupled-cluster and DMC calculations for large  weakly bonded molecules are very difficult to disentangle, as absolute energies can not be compared between the methods. \cite{al2021interactions} 

We have demonstrated that in the thermodynamic limit---that is for many $\mathbf k$ points and virtual bands---ph-AFQMC is 
 superior to a quite efficient Fortran implementation of CCSD(T). 
 This is insofar remarkable, as our own implementation is not yet fully optimized and uses Python. 
 Specifically, CCSD(T) possesses a disadvantageous
 scaling with both the number of $\mathbf k$ points and the number of virtual orbitals. 
 In practice we found that
our ph-AFQMC code scales quadratic with respect to the number of unoccupied orbitals, and cubic with respect
to the number of $\mathbf k$ points. 
This means it will always outpace a CCSD(T) code for sufficiently large systems.

The second important observation is that for the solids investigated here, CCSD(T) and ph-AFQMC agree within chemical accuracy for absolute energies. 
Although we have done the comparison initially for few $\mathbf k$ points and few bands, our tests for diamond clearly show that the differences will not increase as the number of $\mathbf k$ points or bands increases. 
The excellent agreement also means that one can potentially mix and match ph-AFQMC and CCSD(T) in actual investigations and rely on advantages of one or the other method for specific sub-problems.

Last but not least, we have made one somewhat disconcerting observation: our ph-AFQMC correlation energies are generally slightly more negative than the corresponding CCSD(T) energies (but again well within chemical accuracy). 
It is well understood that CCSD(T) often converges from above, so potentially better agreement would be obtained when 
quadruple and pentuple excitation operators are included in the coupled-cluster calculations.
However, ph-AFQMC---being non-variational---also sometimes over-correlates;
so maybe the CCSD(T) values are more accurate after all. 
The available data are clearly not sufficient to draw a final conclusion. Specifically, the reported FCIQMC values hint towards even smaller absolute correlation energies than CCSD(T). 
This we believe to be unlikely: for instance for small molecules, CCSD(T) certainly underestimates the correlation energy  consistently.\cite{Bomble:2005,Zoran:2023} 
Why should this be any different for simple prototypical insulators and semiconductors? 
So reference-type calculations for solids using few $\mathbf k$ points and bands would be very helpful to solve this small but important "riddle". 

\section*{Acknowledgments}

Funding by the Austrian Science Foundation (FWF) within the project P 33440 is gratefully acknowledged. Parts of the presented computational results have been obtained using the Vienna Scientific Cluster (VSC).  

\section*{Author declarations}
\subsection*{Conflict of Interest}
The authors have no conflicts to disclose.

\subsection*{Author Contributions}

Amir Taheridehkordi: Investigation, Methodology, Software, Writing – original draft. Martin Schlipf:  Software, Writing – review \& editing. Zoran Sukurma: Methodology, Writing – review \& editing. Moritz Humer: Writing – review \& editing. Andreas Gr\"uneis: Writing – review \& editing. Georg Kresse: Project administration, Writing – review \& editing.

\section*{Data availability}
The data that support the findings of this study are available within the article. The Python code is available from the first author upon reasonable request.

\appendix

\section{Hamiltonian}\label{appendix:hamil}
Units used in the appendix and throughout the manuscript are Hatree-units. The electronic Born-Oppenheimer Hamiltonian consists of one-body and two-body parts \cite{Born,szabo2012modern}
\begin{eqnarray}\label{E:Hamiltonian}
\hat H = \hat H_1 + \hat H_2. 
\end{eqnarray}
The one-body part is defined as
\begin{eqnarray}\label{E:Hamiltonian_one_body}
\hat H_1 = \sum_{pq} t_{pq} \hat a_{p}^\dagger \hat a_{q}, 
\end{eqnarray}
where $\hat a_{p}$ ($\hat a^\dagger_{p}$) are fermionic annihilation (creation) operators associated with an orthonormal basis $\phi_p$.
The matrix elements are
\begin{eqnarray}\label{E:hoping}
t_{pq}=\int {\rm d} \mathbf r \phi_p^*(\mathbf r) \bigg(-\frac{1}{2} \hat \nabla ^2 -\sum_a \frac{Z_a}{|\mathbf r - \mathbf R_a|}\bigg) \phi_q(\mathbf r),
\end{eqnarray}
where $\mathbf R_a$ and $Z_a$ denote the position and atomic number of the nuclei with label $a$, respectively. 
The two-body part of the Hamiltonian (\ref{E:Hamiltonian}) describes the electron-electron interaction
\begin{eqnarray}\label{E:Hamiltonian_two_body}
\hat H_2 = \frac{1}{2} \sum_{pqrs} \langle pq | rs \rangle \hat a_{p}^\dagger \hat a_{q}^\dagger \hat a_{s} \hat a_{r}
\end{eqnarray}
introducing an abbreviation for the Coulomb integral
\begin{eqnarray}\label{E:two_e_int}
\langle pq | rs \rangle = \int {\rm d} \mathbf r {\rm d} \mathbf r' \phi_p^*(\mathbf r) \phi_q^*(\mathbf r') \frac{1}{|\mathbf r- \mathbf r'|} \phi_r(\mathbf r) \phi_s(\mathbf r').
\end{eqnarray}

Next, we introduce a $\mathbf k$-point mesh to sample the Brillouin zone.
The one-body part is diagonal in the Bloch vector $\mathbf k$
\begin{eqnarray} \label{E:H1_k_rep}
\hat H_1 = \sum_{\mathbf k} \sum_{pq} t_{pq}(\mathbf k) \hat a_{p\mathbf k}^\dagger \hat a_{q\mathbf k}.
\end{eqnarray}
The two-body part has to fulfill momentum conservation \cite{Motta:2019}  
\begin{multline} \label{E:H2_k_rep}
\hat H_2 = \frac{1}{2} \sum_{\mathbf{k}_p + \mathbf {k}_q = \mathbf{k}_r + \mathbf{k}_s}
\sum_{prqs} \langle p \mathbf k_p, q \mathbf k_q | r\mathbf k_r, s\mathbf k_s \rangle \\
\times \hat a_{p\mathbf k_p }^\dagger \hat a_{q \mathbf k_q }^\dagger \hat a_{s \mathbf k_s } \hat a_{r \mathbf k_r},
\end{multline}
which we rewrite introducing the transferred momentum $\mathbf q = \mathbf k_p - \mathbf k_r$
\begin{multline} \label{E:H_2_momentum_transfer}  
  \hat H_2 = \frac{1}{2} \sum_{\mathbf{q} \mathbf{k}_r \mathbf{k}_s} \sum_{prqs}
\langle p \mathbf k_r+\mathbf q, q \mathbf k_s - \mathbf q | r\mathbf k_r, s\mathbf k_s \rangle \\
\times \hat a_{p\mathbf k_r + \mathbf q}^\dagger \hat a_{q \mathbf k_s - \mathbf q}^\dagger \hat a_{s \mathbf k_s} \hat a_{r \mathbf k_r}.
\end{multline}
Using a plane-wave basis set, we express electron-repulsion integrals in the reciprocal space as
\begin{multline} \label{E:ERI_2}
\langle p \mathbf k_r+\mathbf q,q \mathbf k_s - \mathbf q | r\mathbf k_r, s\mathbf k_s \rangle 
= \sum_{\mathbf G} \frac{4\pi}{|\mathbf G - \mathbf q|^2} \\ 
\times \rho_{pr\mathbf k_r} (\mathbf q,\mathbf G) 
\rho^\ast_{sq\mathbf k_s - \mathbf q}(\mathbf q,\mathbf G).
\end{multline}
Here, we introduced the two-orbital density $\rho$ 
\begin{equation}\label{E:def_rho}
 \rho_{pr\mathbf k_r}(\mathbf q, \mathbf G) 
 = \frac{1}{\sqrt{\Omega}} \int {\rm d} \mathbf r {\rm{e}}^{i (\mathbf G  - \mathbf q)\cdot\mathbf r} \phi_{p\mathbf k_r +\mathbf q}^\ast(\mathbf{r}) \phi_{r\mathbf k_r}(\mathbf{r})
\end{equation}
and $\Omega$ is the volume of the system.
The summation over plane waves in Eq.~\eqref{E:ERI_2} is truncated by an energy
cutoff $G^2/2=E_{\rm{cut}}$.

Finally, we commute $\hat a_{r\mathbf k_r}$ to the left in Eq.~(\ref{E:H_2_momentum_transfer})
using the anti-commutation relations between the fermionic operators 
$\{\hat a_r^\dagger,\hat a_p\} = \hat a_r^\dagger \hat a_p + \hat a_p \hat a_r^\dagger = \delta_{rp}$.
This yields the modified one- and two-body operators shown in Eqs.~(\ref{E:H_1}) 
and (\ref{E:H_2}), respectively.
The updated one-body matrix elements are
\begin{eqnarray}\label{E:H_1_tpq}
h_{pq} (\mathbf k) =  t_{pq} (\mathbf k) -\frac{1}{2} \sum_{\mathbf k_r} \sum_{r} \langle p\mathbf k, r\mathbf k_r | r\mathbf k_r, q\mathbf k \rangle.
\end{eqnarray}
We introduce the operators
\begin{equation}\label{E:A_Q_G}
\hat L'_{\mathbf{qG}} = \frac {\sqrt {4\pi}}{|\mathbf G - \mathbf q|} \sum_{\mathbf k_r} \sum_{pr} \rho_{pr \mathbf k_r} (\mathbf q, \mathbf G) \hat a_{p \mathbf k_r + \mathbf q}^\dagger \hat a_{r \mathbf k_r}
\end{equation}
to write the two-body operator in terms of one-body operators:
\begin{eqnarray}\label{E:H_2_appendix}
\hat H_2' = \frac{1}{2} \sum_{\mathbf q \mathbf G}  \hat L'_{\mathbf{qG}}  \hat L'^\dagger_{\mathbf{qG}}.
\end{eqnarray}

\section{VASP interface}\label{appendix:VASP}

For the ph-AFQMC code, we require the one-body matrix $t_{pq}(\mathbf k)$ in Eq.~(\ref{E:H1_k_rep}) and the two-body tensor
\begin{equation}\label{E:L_mat_rep}
  L'_{pr\mathbf k_r\mathbf q\mathbf G} =
  \frac{\sqrt{4\pi}}{|\mathbf G - \mathbf q|} \rho_{pr\mathbf k_r}(\mathbf q,\mathbf G),
\end{equation}
in Eq.~(\ref{E:A_Q_G}). We compute these using  VASP.\cite{Bloechl1994,Kresse1999}
Typically, we use a smaller energy cutoff for the two-body tensor than for the one-body matrix;
this can lead to small inconsistencies in the energy of the HF ground state.
To address this, we pre-calculate the fully self-consistent one-body HF-Hamiltonian matrix $t_{pq}^{\mathrm HF}(\mathbf k)$ within the VASP code. This term includes the kinetic energy, the ion-electron potential, the Hartree potential and the Fock exchange operator.
Details on the VASP implementation are given in Ref.~\onlinecite{Paier:2005}.
For the canonical HF orbitals, the matrix is diagonal and equal to the eigenvalues $t_{pq}^{\mathrm HF}(\mathbf k) = \varepsilon_{p\mathbf k} \delta_{pq}$.
For the ph-AFQMC calculation, we require only the kinetic energy term and the ion-electron potential. 
We compute this term by subtracting the Hartree ($J_{pq}(\mathbf k)$) and Fock contribution ($K_{pq}(\mathbf k)$) calculated from  the two-body tensors (with the
lower energy cutoff, which is equal to the plane-wave energy cutoff) and subtract them to obtain the matrix elements
\begin{equation}
  t_{pq}(\mathbf k) = t_{pq}^{\mathrm HF}(\mathbf k) - J_{pq}(\mathbf k) + K_{pq}(\mathbf k),
\end{equation}
with
\begin{align}\label{E:hartree-mat}
\begin{split}
&J_{pq}(\mathbf k) = 2 \sum_{i\mathbf{k}'\mathbf {G}} \bigg(\sum_{\mathbf q} L'_{\mathbf{qG}} \bigg)_{p\mathbf k, q\mathbf k} \bigg(\sum_{\mathbf q} L'_{\mathbf{qG}}\bigg)^*_{i\mathbf k',i\mathbf k'}, \\
&K_{pq}(\mathbf k) = \sum_{i \mathbf{k}' \mathbf {G}} \bigg(\sum_{\mathbf q} L'_{\mathbf{qG}} \bigg)_{p\mathbf k,i \mathbf k'} \bigg(\sum_{\mathbf q} L'_{\mathbf{qG}}\bigg)^*_{q \mathbf k,i \mathbf k'}.
\end{split}
\end{align}
where index $i$ goes over the occupied states.
This strategy minimized truncation errors that would occur if we just exported the kinetic energy and the electron-ion matrix elements from the VASP code. 
For instance, a self-consistent HF calculation using the one-body Hamiltonian $t_{pq}(\mathbf k)$ and the two body tensors $L'_{\mathbf{qG}}$ yields exactly the same eigenvalues as the preceding VASP calculation.

For the two-orbital tensor defined above, we loop $\mathbf q$ and $\mathbf k_p$ over the mesh
sampling the Brillouin zone.
Each pair corresponds to a $\mathbf k_r = \mathbf k_p - \mathbf q$ conserving the
momentum.
$\mathbf k_r$ may lay outside of the first Brillouin zone;
folding it back with a reciprocal lattice vector $\delta \mathbf G$ introduces a phase $\varphi
= \exp[i \, \delta \mathbf G \cdot \mathbf r]$.
The orbitals $\phi_{r\mathbf k_r}$ consist of a smooth pseudo-orbital $\tilde \phi_{r\mathbf k_r}$
augmented by the difference of atomic orbitals $\phi^{1}_\nu$ and their pseudized counterpart
$\tilde \phi^{1}_\nu$.
Projectors $p_\nu^1$ determine the replaced fraction of the pseudo-orbital
\begin{equation}
  |\phi_{r\mathbf k_r}\rangle = |\tilde \phi_{r\mathbf k_r}\rangle
  + \sum_\nu \bigl( |\phi_\nu^1\rangle - |\tilde \phi_\nu^1\rangle \bigr)
  \langle p_\nu^1 | \tilde \phi_{r\mathbf k_r}\rangle.
\end{equation}
The superscript 1 indicates \emph{one-center} quantities that are only nonzero within
the PAW sphere of one atom.
Non-local operators contain coupling between the pseudo-orbital and the one-center
terms.\cite{Bloechl1994}
Therefore the two-orbital density given by Eq.~(\ref{E:def_rho}) consists
of a pseudo and a one-center contribution 
\begin{equation}
  \rho_{pr\mathbf k_r}(\mathbf q,\mathbf G) =
  \tilde \rho_{pr\mathbf k_r}(\mathbf q,\mathbf G)
  + \rho_{pr\mathbf k_r}^1(\mathbf q,\mathbf G).
\end{equation}
Representing the one-center term would require a very dense real-space grid.
To avoid this one introduces a compensation density $\hat \rho$ that restores the
multi-poles of the all-electron density on the plane-wave grid.\cite{Kresse1999,Paier:2005}
As a result, the one-center density has no Coulomb interaction outside the PAW sphere. We also neglect the contributions inside the PAW sphere in the present work:
\begin{equation}
  \rho_{pr\mathbf k_r}(\mathbf q,\mathbf G) \approx
  \tilde \rho_{pr\mathbf k_r}(\mathbf q,\mathbf G)
  + \hat \rho_{pr\mathbf k_r}(\mathbf q,\mathbf G).
\end{equation}
To make up for the neglect of the terms in the PAW spheres, we use shape restoration. 
Shape restorations allows to accurately restore the all-electron density distribution
inside the PAW spheres even on a coarse plane-wave grid.\cite{shishkin2006implementation,unzog2022x} It is routinely used in VASP for calculations using, e.g., the random phase approximation and sufficiently accurate to obtain highly reliable correlation energy differences.\cite{Moritz:2022} 
Here, we set \texttt{LMAXFOCKAE = 4} to force an accurate treatment for the charge augmentation up to the angular quantum number of 4.
Finally, we combine the two-orbital density with the square of the Coulomb potential and correct
the phase $\varphi$ if necessary.

The one-body matrix $t_{pq}(\mathbf k)$ and the two-body tensor
$L'_{pr\mathbf k_r\mathbf q\mathbf G}$ are each exported in NPY format to
facilitate easy processing in Python.

\section{Time evolution}\label{Sec:B_appendix}
To determine the ground-state wavefunction $|\Phi_0\rangle$ of a system
governed by a Hamiltonian $\hat H$, the imaginary-time 
propagator is applied to the initial wavefunction $|\Psi_{\rm I}\rangle$ in the infinite time limit \cite{Zhang:2003}
\begin{eqnarray}\label{E:AFQMC_core_cont}
|\Phi_0\rangle \propto \lim_{\beta\to\infty} e^{-\beta \hat H} |\Psi_{\rm I}\rangle.
\end{eqnarray}
In practice, one approaches the ground-state wavefunction by repeated application of
the imaginary-time propagator for a small time step $\tau = \frac{\beta}{n}$
\begin{eqnarray}\label{E:AFQMC_core_disc}
\lim_{\beta\to\infty} e^{-\beta \hat H} |\Psi_{\rm I}\rangle = \lim_{n\to\infty} \bigg[{\rm e}^{-\tau \hat H}\bigg]^n |\Psi_{\rm I}\rangle.
\end{eqnarray}
The Hubbard-Stratonovich transformation\cite{Stratonovich:1957,hubbard:1959} translates the two-body part to an integral
of one-body operators
\begin{eqnarray}\label{E:HS_trans}
{\rm e}^{-\tau \hat H} = \int {\rm d} \mathbf x p(\mathbf x) \hat B(\mathbf x) + O(\tau^2).
\end{eqnarray}
$p(\mathbf x)$ is the normal distribution function
\begin{eqnarray}\label{E:norm_dist}
 p(\mathbf x) = (2\pi)^{-N_{g}/2} {\rm e}^{-\frac12 |\mathbf x|^2}, 
\end{eqnarray}
where $N_g$ is the number of components of $\mathbf{x}$.
The propagation operator $\hat B$ is defined in Eq.~(\ref{E:one_body_propag}) in the main text.

\section{Force bias}\label{Sec:fb_appendix} 
The force bias is given by
\begin{eqnarray}\label{E:force_bias_appendix}
f_{\mathbf qg,k}^w = -i \sqrt{\tau} \frac{\langle \Psi_{\rm T} | \hat{\mathfrak L}_{\mathbf qg} | \Psi_{k}^w \rangle}{\langle \Psi_{\rm T} | \Psi_{k}^w \rangle}.
\end{eqnarray}
A creation-annihilation operator pair is
\begin{equation}\label{E:green_func}
    \frac{\langle \Psi_{\rm T} | \hat a_{P}^\dagger \hat a_{R} | \Psi_{k}^w \rangle}{\langle \Psi_{\rm T} | \Psi_{k}^w \rangle} 
    = \Bigl[\Psi_{k}^w\big( \Psi_{\rm T}^\dagger \Psi_{k}^w\big)^{-1} \Psi_{\rm T}^\dagger \Bigr]_{RP}
\end{equation}
in matrix representation. $P$ and $R$ are composite indices for band and momentum, i.e., $R \equiv (r\mathbf k_r)$.
We compute the biorthogonalized orbitals in every step
\begin{equation}
    \Theta_{k}^w = \Psi_{k}^w\big( \Psi_{\rm T}^\dagger \Psi_{k}^w \big)^{-1}.
\end{equation}
The action of the $\hat {\mathfrak L}$ operator onto the trial wavefunction is precomputed
\begin{equation} \label{E:alpha_e_o}
    \alpha_{{\mathbf qg}} = \Psi_{\rm T}^\dagger \mathfrak L_{{\mathbf qg}}~.
\end{equation}
We convolute these two matrices to obtain the force bias for a non-spin-polarized system
\begin{equation}\label{E:force_bias_e_o_fin}
    f_{{\mathbf qg,k}}^w = -2i\sqrt{\tau}{\rm Tr} [\Theta_{k}^w \alpha_{\mathbf{q}g}].
\end{equation}

\section{Local energy}\label{Sec:loc_appendix}
The local energy is 
\begin{eqnarray}\label{E:loc_energy_appendix}
E_{\rm{loc}}(\Psi_{k}^w) = \frac{\langle \Psi_{\rm T}| \hat H_1 + \hat H_2 | \Psi_{k}^w\rangle}{\langle \Psi_{\rm T}| \Psi_{k}^w\rangle}. 
\end{eqnarray}
Similar to the force bias we use Eq.~(\ref{E:green_func}) to evaluate the one-body part in the matrix representation. For a non-spin-polarized system we obtain
\begin{align}\label{E:e_loc1}
\begin{split}
E_1(\Psi_{k}^w) = 2{\rm{Tr}} [\Psi_{\rm T}^\dagger H_1 \Theta_{k}^w].
\end{split}
\end{align}
For the two-body part we consider the generalized Wick's theorem: \cite{wick:1950,balian:1969}
\begin{align}\label{E:wick_theorem}
    \begin{split}
       \frac{\langle \Psi_{\rm T} | \hat a_{P}^\dagger \hat a_{Q}^\dagger \hat a_{S} \hat a_{R} | \Psi^w \rangle}{\langle \Psi_{\rm T} | \Psi^w \rangle}   =  \mathcal G_{PR}^w\mathcal G_{QS}^w -\mathcal G_{PS}^w\mathcal G_{QR}^w.
    \end{split}
\end{align}
Here, $\mathcal G$ is the Green's function
\begin{equation}
    \mathcal G_{PR} ^w \equiv \mathcal G_{PR} (\Psi^w) = \frac{\langle \Psi_{\rm T} | \hat a_{P}^\dagger \hat a_{R} | \Psi^w \rangle}{\langle \Psi_{\rm T} | \Psi^w \rangle} 
\end{equation}
that we know to compute from Eq.~\eqref{E:green_func}.

Thus, the two-body part of the local energy is split into Hartree- and exchange-like terms:
\begin{align}\label{E:e-loc_2}
\begin{split}
     E_2(\Psi_{k}^w) =   E_{\rm H}(\Psi_{k}^w) +  E_{\rm X}(\Psi_{k}^w).
\end{split}
\end{align}
For a non-spin-polarized system:
\begin{align}\label{E:hartree_term}
\begin{split}
E_{\rm H}(\Psi_{k}^w) = 2\sum_{\mathbf{q}}\sum_{g=1}^{N_{\mathbf q g}} {\rm{Tr}}[\Theta_{k}^{w} \alpha_{{\mathbf{q}}{{g}}}] {\rm{Tr}}[\Theta_{k}^{w} \beta_{{\mathbf{q}}{{g}}}], \\
E_{\rm X}(\Psi_{k}^w) = -\sum_{\mathbf{q}} \sum_{g=1}^{N_{\mathbf q g}}{\rm{Tr}}[(\Theta_{k}^{w} \alpha_{{\mathbf{q}}{{g}}}) (\Theta_{k}^{w} \beta_{{\mathbf{q}}{{g}}}) ],
\end{split}
\end{align}
where for each specific $g$ we have 
\begin{align}\label{E:alpha_beta}
\begin{split}
&\beta_{{\mathbf{q}}{{g}}} = \Psi_{\rm T}^\dagger \mathfrak L^\dagger_{{\mathbf qg}},
\end{split}
\end{align}
and $\alpha_{{\mathbf{q}}{\rm{g}}}$ is given by Eq.~(\ref{E:alpha_e_o}).

\bibliography{ref.bib}

\end{document}